\documentstyle[12pt,epsfig,amsfonts]{article}
\newcommand{\beq}{\begin{eqnarray}}
\newcommand{\eeq}{\end{eqnarray}}
\newcommand{\eqn}{\begin{equation}}
\newcommand{\een}{\end{equation}}
\begin{document}
\title{Supersymmetry versus Integrability \\ in two-dimensional Classical Mechanics}
\author{A. Alonso Izquierdo$^{(a,b)}$, M.A. Gonz\'alez Le\'on$^{(b)}$,\\ J. Mateos
Guilarte$^{(c)}$, and M. de la Torre Mayado$^{(c)}$
\\  \\{\normalsize {\it $^{(a)}$ DAMPT}, {\it Cambridge University, U.K.}}\\
{\normalsize {\it $^{(b)}$ Departamento de Matem\'atica Aplicada},
{\it Universidad de Salamanca, SPAIN}}\\{\normalsize {\it $^{(c)}$
Departamento de F\'{\i}sica}, {\it Universidad de Salamanca,
SPAIN}}}
\date{}
\maketitle
\begin{abstract}
Supersymmetric extensions  of Hamilton-Jacobi separable Liouville
mechanical systems with two degrees of freedom are defined. It is
shown that supersymmetry can be implemented in this type of
systems in two independent ways. The structure of the constants of
motion is unveiled and the entanglement between integrability and
supersymmetry is explored.
\end{abstract}

\section{Introduction}

Since the discovery of superstrings \cite{Ram} and non-linear
supersymmetric field theoretical models \cite{Wes} at the
beginning of the seventies in the past century supersymmetry has
become an extremely active area of research in both theoretical
physics and mathematics. In particular, supersymmetric non-Abelian
gauge theories in several dimensions are of broad interest passing
through the phenomenology of elementary particles \cite{Wein} to
differential invariants of four manifolds \cite{Del}.

At a very early stage in the development of this matter, several
researchers started to focus on understanding Bose/Fermi symmetry
in the realms of classical and quantum mechanics of finite
dimensional dynamical systems \cite{Casal}, \cite{Berezin}. In
supersymmetric quantum mechanics, the fermionic variables are
realized as the generators of a Clifford algebra satisfying the
quantization rules: $\{ \theta^j(t),\theta^k(t) \}=i \hbar
\delta^{jk}$. Therefore, they are (Grassmann) anticommuting
variables $\{ \theta^j,\theta^k \}=0$ at the $\hbar=0$ classical
limit \cite{Freund}. The geometry of manifolds including variables
of this kind is described in DeWitt's  book \cite{Witt}. Thus, in
classical supersymmetric theories the configuration space is a
supermanifold, in DeWitt's sense: the dynamical variables take
their values in a Grassman algebra ${\cal B}_L$. Any element of
${\cal B}_L$ is the combination of the $L$ generators
$\xi_{A_k}^{i_r}$, $b=b_0 {\bf 1}+\sum b_{i_1 \cdots i_n}^{A_1
\cdots A_m} \xi_{A_1}^{i_1} \cdots \xi_{A_m}^{i_n}$, where the
coefficients $b_{i_1 \cdots i_n}^{A_1 \cdots A_m}$ are real
numbers. $b_0$ is usually referred to as the body of the
super-number $b$ whereas the sum of the other terms in the
Grassman expansion is accordingly named as the soul of $b$.  The
Lagrangian formalism of classical mechanics can be extended to the
supersymmetric framework and the Hamiltonian formalism can also be
implemented in supersymmetric dynamical systems, see
\cite{Junker,Dirac}. Several simple mechanical models, with
bosonic and fermionic dynamical variables valued in a Grassman
algebra have been investigated by Casalbuoni \cite{Casal}, Berezin
and Marinov \cite{Berezin}, Junker and Matthiesen \cite{Junker2}.
Manton and Heumann have recently improved on these works,
obtaining supersolutions in several interesting models
\cite{Manton1,Manton2,Heumann}

The main theme in the present work is to investigate the interplay
between supersymmetry and integrability in dynamical systems with
two bosonic degrees of freedom. There is a broad class of
two-dimensional classical systems, called Liouville systems, that,
besides being completely integrable, have the stronger property of
being Hamilton-Jacobi-separable, see \cite{Li}. Amongst them rank
some important physical systems: the two-dimensional Kepler and
two Newtonian centers of force problems, the Garnier system
\cite{Gar}, to quote just three. The existence of a second
first-integral in involution with the Hamiltonian guarantees, via
Liouville's theorem, complete integrability. The second invariant
is usually referred as corresponding to hidden symmetries; we
shall show that, via the introduction of generalized momenta,
these invariants can be related with well known constants of
motion. Moreover, the hallmark of Liouville's systems, the
separability of the Hamilton-Jacobi equation, reduces the
analytical solution to independent quadratures in the two degrees
of freedom.

In our attempt to build supersymmetry in systems of the Liouville
type we face three main tasks:
\begin{enumerate}

\item Construction of supersymmetric extensions to be called
superLiouville models.

\item The search for the second invariant in the supersymmetric
framework.

\item To look at the fate of Hamilton-Jacobi separability when
fermionic degrees of freedom are added in a supersymmetric way
\end{enumerate}

There are precedents of the study of constant of motions in
supersymmetric classical mechanics in the literature. Plyushchay
identified invariants in supersymmetric classical and pseudo
classical mechanical models involving one bosonic degree of
freedom in \cite{Pl}. Heumann \cite{Heumann} and Wipf et al
\cite{Wipf} dealt with the supersymmetric version of the
Runge-Lenz vector respectively in the classical and quantum
supersymmetric Coulomb problems. Whether or not the invariants of
a classical system promote invariants in the supersymmetric
extension is a crucial question regarding integrability. There are
also precedents of connecting classical integrable systems with
supersymmetry, see the recent paper \cite{Ioffe} where the
classical limit of SUSY quantum mechanics is used to define two
dimensional integrable systems. Connections between non-linear
supersymmetry and quasi-exactly solvable systems have been pointed
out in the interesting papers \cite{Plyushchay1} and
\cite{Plyushchay2}, where the dynamics of a charged spin ${1\over
2}$ in a magnetic field is described using ideas of SUSY quantum
mechanics.

In supersymmetric models with one bosonic degree of freedom the
Hamiltonian and the supercharges are obvious (non-independent)
first-integrals. The analytic solution of these models can thus be
reduced to quadratures via the Grassman/Manton/Heumann expansion
as in \cite{Manton2}. There also exist some combinations between
the fermionic variables which are conserved, such as was shown in
\cite{Pl}. The situation is more difficult in models with two
bosonic degrees of freedom, where the identification of invariants
is an arduous task even within the purely bosonic framework.
Regarding this latter point, one of most celebrated work is that
of Hietarinta \cite{Hietarinta}: all the integrable systems of
type $H=\frac{1}{2}p_{x_1}^2+\frac{1}{2} p_{x_2}^2+U(x_1,x_2)$ are
analyzed, with $U$ a polynomial in $x_1$ and $x_2$ of degree 5 or
less, and the second invariant is at most of fourth order in $p_x$
and $p_y$. The procedure used is conceptually simple: the
existence of a second invariant $I_2$ in involution with the
Hamiltonian guarantees classical integrability in two-dimensional
systems. A polynomial of any order in the coordinates, but at most
of degree 4 in the  momenta  with arbitrary coefficients, is
proposed as a candidate to become the second invariant $I$.
Accordingly, the Poisson bracket $\{H,I\}_P$ is computed. Systems
where $I$ can be found such that $\{H,I\}_P=0$ are integrable.
This strategy has been followed in other works \cite{Gra,Ra} and
was extended to supersymmetric systems in \cite{AGM}. Here, we
shall apply again Hietarinta's method to two-dimensional
supersymmetric classical mechanics in the search for the second
invariant in superLiouville models. This will be possible because
the Hamiltonian formalism has also been well-defined in the
supersymmetric framework \cite{Dirac}.

Below we offer the conclusions that we have reached with respect
to the three tasks:
\begin{enumerate}

\item There exist ${\cal N}=2$ extended supersymetric versions of
Liouville models. Usually, interactions in supersymmetric theories
are determined from the superpotential. The surprise is that in
superLiouville models one can choose between two different
superpotentials leading to different supersymmetric dynamics. The
reason is that the Hamilton-Jacobi equation theory of the parent
Liouville model admits four different solutions for the Hamilton
characteristic function ( the superpotential ). Two of the
solutions differ from the other two by a global sign that makes no
difference at all. The other two induce different Bose/Fermi
Yukawa couplings.

\item There exist second invariants in other the superLiouville
models. The Bose contribution to the Hamiltonian does not depend
on the choice of the superpotential but the Fermi contribution,
the Yukawa couplings, is different for the non-equivalent
superpotentials. Exactly the same situation occurs with respect to
the second invariants.

\item The superLiouville models are not Hamilton-Jacobi separable.
The Yukawa terms spoil the separability of the two degrees of
freedom.

\end{enumerate}

On the physical side, we mention two applications of
supersymmetric classical systems. First, the structure of the
fermionic contribution to the second invariant naturally shows how
the spin of the superparticle is determined as a quadratic product
of Grassman variables: $s^{12}=i\theta_\alpha^1\theta_\alpha^2$.
This explanation for the fermionic degrees of freedom comes back
to \cite{Berezin} and was given a deep group theoretical meaning
by Azcarraga and Lukierski in Reference \cite{Azca}. Second,
${\cal N}=2$ superLiouville models can be understood as the
dimensional reduction of ${\cal N}=1$ supersymmetric field theory
in (1+1)-dimensions. From this point of view the separatrix (
finite action ) supertrajectories are seen as the BPS superkinks
of the field theory ( supersymmetric BPS domain walls in
(3+1)-dimensions) . In References \cite{We1,We2,We3} we have
approached the problem from this angle of attack.

The organization of the paper is as follows: In Section \S 2 we
describe ${\cal N}=2$ supersymmetric classical mechanics, both in
Euclidean and Riemannian 2D manifolds; the notations and
conventions are introduced also in this Section. Section \S 3 is
devoted to defining ${\cal N}=2$ superLiouville systems after a
rapid summary of the properties of the parent Liouville models. In
Section \S 4, the second ( and other ) invariants of the
superLiouville models are identified, following the above
mentioned Hietarinta strategy. Finally, in Section \S 5 a
procedure is outlined to generalize the Hamilton-Jacobi method to
the supersymmetric framework. It is shown how to search for the
supersolutions of the system in a layer-by-layer resolution.

\section{${\cal N}=2$ supersymmetric classical mechanics}

In this Section we introduce ${\cal N}=2$ super-symmetric
classical mechanics, which can also be described as a dimensional
reduction of ${\cal N}=1$ supersymmetric (1+1)-dimensional field
theory. We restrict ourselves to models with two bosonic degrees
of freedom.

There is in the whole formalism an underlying Grassman algebra
${\cal B}_L$ with $L$ odd generators $\xi_A$ such that
$\xi_A\xi_B=-\xi_B\xi_A, A,B=1,2,\cdots , L$, see References
\cite{Witt} and \cite{Manton1}. The Grassman algebra is the direct
sum of even and odd sub-algebras: ${\cal B}_L={\cal B}_L^e+{\cal
B}_L^o$.

\subsection{${\cal N}=2$ super-time and configuration super-space}

The evolution is characterized by the ${\cal N}=2$ super-time
which is the ${\Bbb R}^{1|2}$ super-manifold in the terminology of
Reference \cite{Deligne}. A given \lq\lq super-instant"
$(t,\tau^1,\tau^2)$ is determined by the even, $t\in {\cal
B}_L^e$, and odd, $\tau^\alpha\in {\cal B}_L^o, \alpha=1,2$,
parameters, satisfying the commutation/anti-commutation relations:
\[
[t,\tau^\alpha]=0 \qquad , \qquad
\{\tau^\alpha,\tau^\beta\}=0\qquad .
\]
The ${\cal N}=2$ configuration super-space is the ${\cal C}={\Bbb
R}^{2|4}$ super-manifold. A super-point in ${\cal C}$ is
determined by the coordinates
$(x^j,\theta^j_\alpha),\,j=1,2,\,\alpha=1,2$, satisfying the
commutation/anti-commutation rules:
\[
[x^j,\theta^k_\alpha]=0\qquad , \qquad \{ \theta_\alpha^j ,
\theta_\beta^k\}=0 \qquad , \qquad
\{\theta^j_\alpha,\tau^\beta\}=0 \qquad .
\]
Thus, $(x^1,x^2,\theta^1_1,\theta^1_2,\theta^2_1,\theta^2_2)\in
{\cal B}_L^e\times {\cal B}_L^e\times {\cal B}_L^o\times {\cal
B}_L^o\times {\cal B}_L^o\times {\cal B}_L^o$ and a \lq\lq
super-path"
\[
X^j(t,\tau^1,\tau^2):{\Bbb R}^{1|2}\longrightarrow {\Bbb R}^{2|4}
\]
in ${\cal C}$ is given in terms of its components as:
\[
X^j(t,\tau^1,\tau^2)=x^j(t)+\theta^j_\alpha(t)\tau^\alpha+iF^j(t)\tau^1\tau^2
\quad .
\]
The $F^j(t)$ components of the super-path are needed to match the
number of \lq\lq bosonic" (even), $x^j,F^j$, and \lq\lq fermionic"
(odd), $\theta^j_\alpha$, degrees of freedom.

Besides the time-translation invariance generated by the
Hamiltonian operator ${\tilde H}=-i\partial_t$, we seek a
super-dynamics that is also invariant under the two left
super-time-translations:

\[
\begin{array}{llll}
\mbox{supersymmetry 1:} & \tau^1 \rightarrow \tau^1-\varepsilon^1 &
\tau^2 \rightarrow  \tau^2 & t \rightarrow t-i \tau^1
\varepsilon^1 \\ \mbox{supersymmetry 2:} & \tau^1 \rightarrow
\tau^1 & \tau^2 \rightarrow  \tau^2 -\varepsilon^2 & t \rightarrow
t-i \tau^2 \varepsilon^2
\end{array}
\]
where $\varepsilon^\alpha$ is an infinitesimal odd parameter. The
generators of these transformations are the super-charges,
\[
{\tilde Q}_1=i\tau^1\partial_t-\partial_1 \qquad , \qquad {\tilde
Q}_2=i\tau^2\partial_t-\partial_2 \qquad , \qquad
\partial_\alpha=\frac{\partial}{\partial\tau^\alpha}\qquad ,
\]
that close the ${\cal N}=2$ super-algebra
\begin{equation}
\{{\tilde Q}_\alpha ,{\tilde
Q}_\beta\}=2\delta_{\alpha\beta}{\tilde H} \qquad , \qquad
[{\tilde Q}_\alpha,{\tilde H}]=0 \qquad , \qquad [{\tilde
H},{\tilde H}]=0 \label{eq:salg}
\end{equation}
with the Hamiltonian. The action of the super-charges on a
super-path expressed on the component paths is:
\[ \begin{array}{cc}
\mbox{{\it Supersymmetry 1}  } & \mbox{{\it Supersymmetry 2} }
\\[0.2cm] \delta_1 X^j=\varepsilon {\tilde Q}_1 X^j \Rightarrow \left\{
\begin{array}{l} \delta_1 x^j= \varepsilon \theta_1^j \\ \delta_1
\theta_1^j=i \varepsilon \dot{x}^j \\ \delta_1 \theta_2^j=-i
\varepsilon F^j \\ \delta_1 F^j=-\varepsilon \dot{\theta}_2^j
\end{array} \right. & \delta_2 X^j=\varepsilon {\tilde Q}_2 X^j
\Rightarrow \left\{ \begin{array}{l} \delta_2 x^j= \varepsilon
\theta_2^j \\ \delta_2 \theta_1^j=i \varepsilon F^j \\ \delta_2
\theta_2^j=i \varepsilon \dot{x}^j \\ \delta_2 F^j=\varepsilon
\dot{\theta}_1^j \end{array} \right.
\end{array}
\]

The generators $D_\alpha=i\tau^\alpha\partial_t+\partial_\alpha$
of right super-time-translations are usually called covariant
derivatives because $\{{\tilde Q}_\alpha , D_\beta\}=0$ and:
\[
\delta_\beta D_\alpha X^j=\varepsilon {\tilde Q}_\beta D_\alpha
X^j=D_\alpha \delta_\beta X^j \qquad .
\]
From the free super-Lagrangian ${\cal L}_0$,
\begin{eqnarray}
{\cal
L}_0[x^j,\theta^j_\alpha,F^j]&=&\frac{1}{4}\varepsilon^{\alpha\beta}
D_\alpha X^jD_\beta X^j\nonumber\\ &=&
\frac{1}{4}\theta^j_1\theta^j_2-\frac{i}{2}\tau^1({\dot
x}^j\theta^j_2+\theta^j_1F^j)\nonumber\\&&+\frac{i}{2}\tau^2({\dot
x}^j\theta^j_1-\theta^j_2F^j)+\tau^2\tau^1\left(\frac{1}{2}{\dot
x}^j{\dot
x}^j+\frac{i}{2}\theta^j_\alpha{\dot\theta}^j_\alpha+\frac{1}{2}F^jF^j\right)\label{eq:flag}
\end{eqnarray}
and the \lq\lq super-potential", a function
$W[X^j(t,\tau^1,\tau^2)]$ of the super-path,
\begin{equation}
W[X^j(t,\tau^1,\tau^2)]=W[x^j(t)]-\tau^\alpha\theta^j_\alpha\frac{\partial
W}{\partial x^j}+\tau^1\tau^2\left(iF^j\frac{\partial W}{\partial
x^j}-\frac{\partial^2 W}{\partial x^j\partial
x^k}\theta^j_1\theta^k_2\right) \qquad , \label{eq:cosup}
\end{equation}
the \lq\lq super-action" is built:
\begin{equation}
S=\int \, dtd\tau^1d\tau^2 \, \left[
\frac{1}{4}\varepsilon^{\alpha\beta}D_\alpha X^jD_\beta
X^j+iW[X^j]\right] \qquad , \label{eq:susac}
\end{equation}
which is invariant under the two left super-time-translations.
Here, $\varepsilon^{\alpha\beta}$ is the completely antisymmetric
symbol:
$\varepsilon^{12}=-\varepsilon^{21}=1,\varepsilon^{\alpha\alpha}=0
$. To check that the transformations generated by ${\tilde
Q}_\alpha$ are \lq\lq super-symmetries" of $S$ is not difficult:
\[
\delta_\alpha {\cal L}_0=\varepsilon {\tilde Q}_\alpha {\cal
L}_0=\varepsilon (i\tau^\alpha\partial_t-\partial_\alpha ){\cal
L}_0 \qquad , \qquad  \delta_\alpha W=\varepsilon {\tilde
Q}_\alpha W=\varepsilon (i\tau^\alpha\partial_t-\partial_\alpha )W
\quad .
\]
Both $i\varepsilon\tau^\alpha\partial_t{\cal L}_0$ and
$i\varepsilon\tau^\alpha\partial_tW$ are exact time derivatives
and as such do not contribute to variations of $S$.
$\partial_\alpha{\cal L}_0$ and $\partial_\alpha W$ are at most
linear in $\tau^\beta$. The Berezin integration measure on odd
Grassman variables,
\[
\int d\tau^\alpha =0 \qquad , \qquad \int d\tau^\alpha \,
\tau^\beta= \delta^{\alpha\beta} \quad ,
\]
tells us that $\int d\tau^1 d\tau^2 \, \partial_\alpha{\cal
L}_0=\int d\tau^1 d\tau^2 \, \partial_\alpha W=0$.
\subsection{Lagrangian and Hamiltonian formalism in Euclidean ${\Bbb R}^{2|4}$}
Berezin integration in $S$, (\ref{eq:susac}), plus use of the
constraint equations $F^j=\frac{\partial W}{\partial x^j}$  to
eliminate the auxiliary bosonic variables lead us to the
supersymmetric Lagrangian:
\begin{equation}
L=\frac{1}{2} \dot{x}^j \dot{x}^j + \frac{i}{2} \theta_a^j
\dot\theta_a^j -\frac{1}{2} \frac{\partial W}{\partial x^j}
\frac{\partial W}{\partial x^j} -i \frac{\partial^2 W}{\partial
x^j \partial x^k} \theta_1^j \theta_2^k\label{eq:superlag}
\end{equation}
The Lagrangian is defined on even elements of ${\cal C}$. Besides
the natural Lagrangian on the bosonic degrees of freedom with a
positive semi-definite potential term, the Grassmannian kinetic
energy and a Yukawa coupling between bosonic and fermionic degrees
of freedom enter to guarantee supersymmetry. The necessary and
sufficient condition for extending classical mechanical systems to
the supersymmetric framework is therefore:
\begin{equation}
U(x_1,x_2)=\frac{1}{2} \frac{\partial W}{\partial x^j}
\frac{\partial W}{\partial x^j}\qquad . \label{eq:potsup}
\end{equation}
The potential energy $U(x^j)$ is equal to the square of the norm
of the gradient of the superpotential.

The Euler-Lagrange equations are:
\begin{equation}
\ddot{x}^k+\frac{\partial U}{\partial x^k}+ i \frac{\partial^3
W}{\partial x^j \partial x^l \partial x^k} \theta_1^j \theta_2^l=0
\hspace{1cm} \dot{\theta}_1^i= \frac{\partial^2 W}{\partial x^i
\partial x^j} \theta_2^j \hspace{1cm} \dot{\theta}_2^i=
-\frac{\partial^2 W}{\partial x^j \partial x^i} \theta_1^j \qquad
.. \label{eq:eleq}
\end{equation}
Looking at the first formula in (\ref{eq:eleq}), we notice that
even though the bosonic variables were real ordinary magnitudes at
the initial time, the evolution of the system would convert them
into Grassmannian even variables.

N$\ddot{\rm o}$ether's theorem dictates that the Hamiltonian
functions associated to the vector fields ${\tilde H},{\tilde
Q}_1,{\tilde Q}_2$ are respectively:
\begin{eqnarray*}
H&=&\frac{1}{2}\dot{x}^j\dot{x}^j+\frac{1}{2}\frac{\partial
W}{\partial x^j}\frac{\partial W}{\partial x^j}+i \frac{\partial^2
W}{\partial x^j \partial x^k} \theta_1^j \theta_2^k\\
Q_1&=&\dot{x}^j \theta_1^j - \frac{\partial W}{\partial x^j}
\theta_2^j \hspace{0.5cm} \hspace{0.5cm} Q_2=\dot{x}^j \theta_2^j
+ \frac{\partial W}{\partial x^j} \theta_1^j
\end{eqnarray*}
$H$, $Q_1$ and $Q_2$ are thus first-integrals for the system of
ODE (\ref{eq:eleq}).

We shall now briefly discuss the Hamiltonian formalism
\cite{Junker}, in order to describe how $H$, $Q_1$ and $Q_2$
induce the flows associated to the time- and
super-time-translations in the co-tangent bundle to the
configuration super-space. The usual definition of generalized
momentum $p_j=\frac{\partial L}{\partial\dot{x}^j}$ is extended to
the Grassmannian variables as follows:
\[
\pi_{\theta_\alpha^j}=L\frac{\stackrel{\leftarrow}{\partial}}{\partial
\dot{\theta}_\alpha^j}=\frac{i}{2} \theta_\alpha^j\qquad .
\]
There is a 12-dimensional phase space $T^* {\cal C}$ with local
coordinates $(x^j,\theta_\alpha^j,p_j,\pi_{\theta_\alpha^j})$.
Note, however, the dependence of the fermionic generalized momenta
on the Grassman coordinates, coming from the fact that the
Grassman kinetic energy is of first-order in time derivatives. The
associated four second-class constraints are enforced through
Grassmann Lagrange multipliers $\lambda_a^j$,\cite{Dirac}:
\[
H_T=\frac{1}{2} p_j p_j+ \frac{1}{2} \frac{\partial W }{\partial
x^j} \frac{\partial W }{\partial x^j}+i \frac{\partial^2
W}{\partial x^j \partial x^k} \theta_1^j \theta_2^k -
(\pi_{\theta_a^j}-\frac{i}{2} \theta_a^j) \lambda_a^j
\]
The motion equations are of the canonical form:
\[
\dot{x}^j=\frac{\partial H_T}{\partial p_j} \hspace{1cm}
\dot{p}_j=-\frac{\partial H_T}{\partial x^j} \hspace{1cm}
\dot{\theta}_\alpha^j= \frac{\partial H_T}{\partial
\pi_{\theta_\alpha^i}} \hspace{1cm} \dot{\pi}_{\theta_\alpha^j}=
\frac{\partial H_T}{\partial \theta_\alpha^j}
\]
Note should be taken of the difference in sign between the bosonic
and fermionic canonical equations. Using the constraint equations
$\frac{\partial H_T}{\partial \lambda^j_\alpha}=0$, we write
\[
H_T=\frac{1}{2} p_j p_j+\frac{1}{2} \frac{\partial W}{\partial
x^j}\frac{\partial W}{\partial x^j}-\frac{\partial^2 W }{\partial
x^j \partial x^k} (\pi_{\theta_2^k}
\theta_1^j-\pi_{\theta_1^j}\theta_2^k)
\]
to rule the right Hamiltonian flow in the phase space.

The Poisson brackets for two generic functions $F$ and $G$ are
also generalized to the Grassman variables:
\[
\{F,G\}_P=\frac{\partial F}{\partial p_j} \frac{\partial
G}{\partial q^j} - \frac{\partial F}{\partial q^j} \frac{\partial
G }{\partial p_j}+ i F \frac{\stackrel{\leftarrow}{\partial}}{\partial \theta_\alpha^j}
\frac{\stackrel{\rightarrow}{\partial}}{\partial \theta_\alpha^j} G - \frac{1}{2} F
\frac{\stackrel{\leftarrow}{\partial}}{\partial \pi_{\theta_\alpha^j}}
\frac{\stackrel{\rightarrow}{\partial}}{\partial \theta_\alpha^j} G - \frac{1}{2} F
\frac{\stackrel{\leftarrow}{\partial}}{\partial \theta_\alpha^j} \frac{\stackrel{\rightarrow}{\partial}}{\partial
\pi_{\theta_\alpha^j}} G- \frac{i}{4} F \frac{\stackrel{\leftarrow}{\partial}}{\partial
\pi_{\theta_\alpha^j}} \frac{\stackrel{\rightarrow}{\partial}}{\partial
\pi_{\theta_\alpha^j}} G
\]
and the canonical equations are of the form
\[
\frac{df}{dt}=\{H_T,f\} \hspace{1cm}
\]
for any $f=x^j$,$\theta_\alpha^j$,$p_j$,$\pi_{\theta_\alpha^j}$.
In general, the time dependence of any observable $F$ is
determined by the Poisson structure: $\frac{dF}{dt}=\{H_T,F\}_P$.
Therefore, the constants of motion, i.e. the invariants, are the
physical observables complying with the relationship
$\{H_T,F\}_P=0$.

In practical terms, it is better to work on the reduced (eight
dimensional) phase space. The reduced Hamiltonian is
\[
H=\frac{1}{2} p_j p_j + \frac{1}{2} \frac{\partial W}{\partial
x^j}\frac{\partial W}{\partial x^j} + i W_{jk} \theta^j_1
\theta^k_2 \qquad , \qquad  W_{jk}=\frac{\partial^2 W }{\partial
x^j \partial x^k} \qquad ,
\]
whereas the reduced Poisson brackets
\[
\{F,G\}_P=\frac{\partial F}{\partial p_j} \frac{\partial
G}{\partial q^j} - \frac{\partial F}{\partial q^j} \frac{\partial
G }{\partial p_j}+ i F \frac{\partial}{\partial \theta_\alpha^j}
\frac{\partial}{\partial \theta_\alpha^j} G
\]
are obtained from  the following Poisson structure:
\[
\{p_j,x^k\}_P=\delta_j^k \hspace{1cm}
\{x^j,x^k\}_P=\{p_j,p_k\}_P=0 \hspace{1cm} \{\theta_\alpha^j,
\theta_\beta^k\}_P=i \delta^{jk} \delta_{\alpha\beta}\qquad .
\]
Also, the canonical equations and the invariant observables must
be referred to the reduced Hamiltonian $H$. The most remarkable feature of the super-charges
\[
Q_1=p_j\theta^j_1-\frac{\partial W}{\partial x^j}\theta^j_2\qquad
\qquad Q_2=p_j\theta^j_2+\frac{\partial W}{\partial x^j}\theta^j_1
\]
is seen through the Poisson structure:
\begin{eqnarray}
\{ Q_1,Q_1 \}_P=2H \hspace{1cm}&& \hspace{1cm}\{ Q_2,Q_2 \}_P=2H
\nonumber
\\\{Q_\alpha , H\}_P=0 \hspace{1cm}&&\hspace{1cm} \{ Q_1,Q_2
\}_P=-ip_j\frac{\partial W}{\partial x^j}\quad . \label{eq:exsus}
\end{eqnarray}
The Hamiltonian functions $Q_\alpha$ and $H$ close a central
extension of the ${\cal N}=2$ SUSY algebra (\ref{eq:salg}) by a
topological term: $Z=-ip_j\frac{\partial W}{\partial x^j}$ is a
total derivative with physical implications only if non-trivial
boundary conditions or a non-trivial topology of the configuration
superspace are considered. The flow generated by $Q_\alpha$ in the
co-tangent bundle to the configuration super-space,
\[
\begin{array}{cc}
\mbox{{\it Supersymmetry 1}  } & \mbox{{\it Supersymmetry 2} }
\\[0.2cm]
\begin{array}{l} \delta_1 x^j=\varepsilon \{Q_1,x^j\}_P=\varepsilon \theta_1^j \\ \delta_1
\theta_1^j=\varepsilon \{Q_1,x^j\}_P=i \varepsilon p_j \\ \delta_1
\theta_2^j=\varepsilon \{Q_1,x^j\}_P=-i \varepsilon \frac{\partial
W}{\partial x^j}
\end{array}  &  \begin{array}{l} \delta_2 x^j= \varepsilon \{Q_2,x^j\}_P=\varepsilon \theta_2^j
\\ \delta_2 \theta_1^j=\varepsilon \{Q_2,x^j\}_P=i \varepsilon \frac{\partial W}{\partial x^j}
\\ \delta_2 \theta_2^j=\varepsilon \{Q_2,x^j\}_P=i
\varepsilon p_j \end{array}
\end{array} \qquad ,
\]
represents the two odd super-time translations. Thus, there is a
bosonic invariant, the Hamiltonian $H$ itself, due to invariance
of the theory with respect to even super-time translations. There
are also two fermionic constants of motion, the super-charges
$Q_1$ and $Q_2$ - their Poisson bracket with $H$ is zero- showing
the invariance of the system with respect odd super-time
translations.

As in every ${\cal N}=2$ super-symmetric theory, there is an
$R$-symmetry with respect to rotations of the components of the
fermionic variables. In our system, $\theta^j_\alpha$ can be
understood as Grassman Majorana spinors and one checks that the
reduced Hamiltonian is invariant under the $R$ transformation:
\[
\tilde{\theta}^j_1={\rm cos}\omega\theta^j_1+{\rm
sin}\omega\theta^j_2\qquad , \qquad \tilde{\theta}^j_2=-{\rm
sin}\omega\theta^j_1+{\rm cos}\omega\theta^j_2 \quad .
\]
The infinitesimal generator of this fermionic symmetry is the,
quadratic in the odd coordinates but bosonic, invariant
$S_2=i\theta^j_1\theta^j_2$, see Reference \cite{Manton1}:
\[
\{S_2,\theta^j_1\}_P=\theta^j_2 \qquad , \qquad
\{S_2,\theta^j_2\}_P=-\theta^j_1 \qquad .
\]
It was pointed out by Plyushchay, \cite{Pl}, that there is another
bosonic invariant, $S_3=\prod_{i=1}^N\theta_1^i \theta_2^i$ when
the number of degrees of freedom is $N$. If $N=2$ ,
$S_3=\theta_1^1 \theta_2^1 \theta_1^2 \theta_2^2=-\frac{1}{2}S
_2^2$ and also comes from the $R$-symmetry. The main goal of this
work is to study what happens with super-symmetric extensions in
integrable bosonic dynamical systems, where more invariants than
the Hamiltonian exist.

\subsection{Supersymmetric classical mechanics on Riemannian manifolds}

If the bosonic piece of the configuration super-space is a general
Riemannian manifold $M^2$ equipped with a metric tensor $g_{ij}$,
the Grassman variables $\vartheta_\alpha^j$ under a local change
of coordinates $x^j \rightarrow \tilde{x}^j$ changes as:
$\tilde{\vartheta}^j_\alpha=\frac{\partial \tilde{x}^j}{\partial
x^k}\vartheta_\alpha^k$.

If the \lq\lq body" of the configuration super-space is a two
dimensinal Riemannian manifold the ${\cal N}=2$ super-symmetric
action reads:
\begin{equation}
S=\int dt \, d \tau^1 \, d\tau^2 \left\{\frac{1}{4} g_{jk}(X^l)
\epsilon^{\alpha \beta} D_\alpha X^j D_\beta X^k +i
W[X^j]\right\}\qquad .\label{eq:acc10}
\end{equation}
The expansion of the super-path in the super-time
\[
X^j[t,\tau^1,\tau^2]=x^j(t)+\vartheta^j_\alpha\tau^\alpha+iF^j(t)\tau^1\tau^2
\]
the metric tensor,
\[
g_{jk}(X^l)=g_{jk}(x^l)+\frac{\partial g_{jk}}{\partial x^l}
\vartheta^l_\alpha \tau^\alpha+\tau^1 \tau^2 \left( i
\frac{\partial g_{jk}}{\partial x^l} F^l-\frac{\partial^2
g_{jk}}{\partial x^r\partial x^s} \vartheta_1^r \vartheta_2^s
\right)
\]
and the super-potential,
\begin{eqnarray*}
W[X^j(t,\tau^1,\tau^2)]&=&W[x^j(t)]-\tau^\alpha\theta^j_\alpha
W_{;j}+\tau^1\tau^2\left(iF^j
W_{;j}-W_{;j;k}\theta^j_1\theta^k_2\right)\\W_{;j}= \frac{\partial
W}{\partial x^j}\qquad &,& \qquad  W_{;j;k}=\frac{\partial^2
W}{\partial x^j
\partial x^k}- \Gamma^l_{jk} \frac{\partial W}{\partial x^l}
\end{eqnarray*}
are richer than for flat manifolds. $\Gamma^l_{jk}$ are the
Christoffel symbols.

Berezin integration and use of the constraint equations for the
auxiliary fields leads to the super-symmetric action:
\begin{equation}
S=\! \int \!\! dt \left\{ \frac{1}{2} g_{jk} \dot{x}^j \dot{x}^k
\! +\frac{i}{2} g_{jk} \vartheta_\alpha^j D_t \vartheta_\alpha^k
\! +\frac{1}{4} R_{jkln} \vartheta_1^j \vartheta_2^l \vartheta_1^k
\vartheta_2^n -\frac{1}{2} g^{jk} \frac{\partial W}{\partial x^j}
\frac{\partial W}{\partial x^k} -i W_{;j;k} \vartheta_1^j
\vartheta_2^k  \right\}\qquad , \label{eq:lagsumet}
\end{equation}
where
\[
D_t \vartheta_\alpha^j=\dot{\vartheta}_\alpha^j+\Gamma^j_{sr}
\dot{x}^r \vartheta_\alpha^s
\]
is the covariant derivative for the Grassman variables,
$R_{jkln}$ is the curvature tensor of the metric, and

\begin{equation} U(x^j) = {1\over 2} g^{jk} {\partial W \over \partial
x^j}{\partial W \over \partial x^k}. \label{eq:spor}
\end{equation}

N$\ddot{\rm o}$ther's theorem applied to the invariance of the
action with respect to the transformation,
\[
\delta_1 x^j= \varepsilon \vartheta_1^j \hspace{0.5cm} ,
\hspace{0.5cm} \delta_1 \vartheta_1^j= i \varepsilon \dot{x}^j
\hspace{0.5cm} , \hspace{0.5cm} \delta_1 \vartheta_2^j=-i
\varepsilon \left(g^{jk} \frac{\partial W}{\partial x^k}-i\Gamma^j_{kl}\vartheta_1^j\vartheta_2^k\right)\quad ,
\]
gives the conserved super-charge:
\[
Q_1=g_{jk} \dot{x}^j \vartheta_1^k-\frac{\partial W}{\partial x^j}
\vartheta_2^j \qquad .
\]
Invariance with respect to
\[
\delta_2 x^j= \varepsilon \vartheta_2^j \hspace{0.5cm} ,
\hspace{0.5cm} \delta_2 \vartheta_1^j= i \varepsilon
\left(g^{jk} \frac{\partial W}{\partial x^k}-i\Gamma^j_{kl}\vartheta_1^j\vartheta_2^k\right) \hspace{0.5cm} , \hspace{0.5cm}
\delta_2 \vartheta_2^j= i \varepsilon \dot{x}^j
\]
leads to the second conserved super-charge:
\[
Q_2=g_{jk} \dot{x}^j \vartheta_2^k+\frac{\partial W}{\partial x^j}
\vartheta_1^j \qquad .
\]
The Hamiltonian is:
\begin{equation}
H=\frac{1}{2} g_{jk} \dot{x}^j \dot{x}^k+\frac{1}{2} g^{jk}
\frac{\partial W}{\partial x^j}\frac{\partial W}{\partial x^k}+i
W_{;j;k} \theta_1^j \theta_2^k \label{eq:hamsumet}
\end{equation}

\section{Integrability versus supersymmetry:
from Liouville to SuperLiouville Models}

An $N$-dimensional Hamiltonian system is said to be completely
integrable in the sense of Arnold-Liouville if there exist $N$
integrals of motion, $I_1, \cdots ,I_N$, which are in involution;
i.e.,
\[
\{I_j,I_k\}_P=0 \qquad , j,k=1,2, \cdots , N ,
\]
see e.g. \cite{Tabor}. In practical terms, explicit integration of
the motion equations is more accessible if the Hamilton-Jacobi
equation is separable in some appropriate system of coordinates.
In such a case, a complete solution of the Hamilton-Jacobi
equation is available that, in turn, provides explicit formulas
for the trajectories via the Hamilton-Jacobi method. Choosing
$I_1$ as the Hamiltonian, the time-independent Hamilton-Jacobi
equation for zero energy $I_1(p_j,x^j)=i_1=0$ and $V=-U$ is no
more than the PDE (\ref{eq:spor}). Thus, the body of the
superpotential is the Hamilton's characteristic function for a
natural dynamical system with potential $V=-U$:
$S(x^j,t)=W(x^j)-i_1t$. A supersymmetric mechanical system is
built with a solution of the associated Hamilton-Jacobi equation
from the start.

For a special type of completely integrable system, termed as
Hamilton-Jacobi-separable, the Hamilton-Jacobi equation
(\ref{eq:spor}) becomes equivalent to a system of $N$ non-coupled
ODE's. Liouville systems \cite{Li} are $N=2$
Hamilton-Jacobi-separable systems of the form: $L=\frac{1}{2}
(g_{11}(x^1,x^2) \dot{x}^1 \dot{x}^1+g_{22}(x^1,x^2) \dot{x}^2
\dot{x}^2)-V(x^1,x^2)$. The complete solution of (\ref{eq:spor})
consists of $2^N$ independent solutions coming from the
combinations of the solutions of $N$ one-dimensional problems. In
this Section we shall describe the different $2^N$ supersymmetric
extensions of classical Hamilton-Jacobi separable models obtained
from the distinct $2^N$ Hamilton's characteristic functions, in
the special case of Liouville models \cite{Li}.

\subsection{Liouville models}
Liouville models are two-dimensional completely integrable natural
systems with dynamics governed by Lagrangians of the form:
\[
L=\frac{1}{2} (g_{11}(x^1,x^2) \dot{x}^1 \dot{x}^1+g_{22}(x^1,x^2)
\dot{x}^2\dot{x}^2)-V(x^1,x^2)\qquad .
\]
The key property enjoyed by this class of systems is that the
Hamilton-Jacobi equation for them is separable using
two-dimensional elliptic, polar, parabolic or Cartesian systems of
coordinates. There are four types classified by the kind of system
of coordinates suitable for solving the HJ equation and the two
classical invariants in involution are well known for each
Liouville system. We briefly describe the four possibilities:
\begin{itemize}
\item {\bf Liouville Models of Type I:} Let us
consider the map $\xi: {\Bbb R}^2\longrightarrow {\Bbb D}^2$,
where ${\Bbb D}^2$ is an open sub-set of ${\Bbb R}^2$ with
coordinates $(u,v)$, and let $\xi^{-1}: {\Bbb D}^2\longrightarrow
{\Bbb R}^2$ be the inverse map:
\[
(x^1,x^2)=\xi^{-1}(u,v)=\left( \frac{1}{c}uv, \pm \frac{1}{c}
\sqrt{(u^2-c^2)(c^2-v^2)}\right)
\]
\[
\xi(x^1,x^2)=(u,v)
\]
\begin{eqnarray*}
u&=&{1\over
2}\left(\sqrt{(x^1+c)^2+x^2x^2}+\sqrt{(x^1-c)^2+x^2x^2}\right)
 \\ v&=&\pm{1\over
2}\left(\sqrt{(x^1+c)^2+x^2x^2}-\sqrt{(x^1-c)^2+x^2x^2}\right)
\end{eqnarray*}
The $u,v$ variables are the elliptic coordinates of the bosonic
system: $u \in [c,\infty)$, $v \in [-c, c]$ and ${\Bbb D}^2$ is
the closure of the infinite strip: $\bar{\Bbb
D}^2=[c,\infty)\times [-c, c]$. Let us assume the notation $\xi^*$
for the induced map in the functions on ${\Bbb R}^2$, i.e. $\xi^*
U (x^1,x^2)=U(\xi(x^1,x^2))\equiv U(u,v)$, so we will write $U$
for $U(x^1,x^2)$ and $\xi^* U$ for $U(u,v)$ and a similar
convention will be used for the functions in the phase and
co-phase spaces.

In the new variables the Lagrangian of a Liouville model of Type I
is constrained to be of the following form:
\begin{equation}
\xi^* L = \displaystyle\frac{1}{2}\,
\displaystyle\frac{u^2-v^2}{u^2-c^2} \, \dot{u}\, \dot{u}+
\displaystyle\frac{1}{2} \, \displaystyle\frac{u^2-v^2}{c^2-v^2}
\, \dot{v}\, \dot{v} - \displaystyle\frac{u^2-c^2}{u^2-v^2} \,
f(u)-\displaystyle\frac{c^2-v^2}{u^2-v^2}
 \, g(v)
\label{eq:Liou1}
\end{equation}
where $f(u)$ and $g(v)$ are arbitrary functions. Observe that,
apart from a common factor, the contribution to the Lagrangian of
the $u$ and $v$ variables splits completely. The Hamilton-Jacobi
equation for zero energy and $V=-U$, formula (\ref{eq:spor}),
written in elliptic coordinates reads:
\begin{equation}
\xi^* U= \displaystyle\frac{u^2-c^2}{u^2-v^2} \,
f(u)+\displaystyle\frac{c^2-v^2}{u^2-v^2} \, g(v) =\frac{1}{2}
\displaystyle\frac{u^2-c^2}{u^2-v^2} \left(\frac{d F}{du}
\right)^2 +\frac{1}{2} \displaystyle\frac{c^2-v^2}{u^2-v^2}
\left(\frac{d G}{dv} \right)^2 \label{eq:ehj}\quad .
\end{equation}
assuming separability: $\xi^* W=F(u)+G(v)\Rightarrow
\frac{\partial^2 \xi^*W}{\partial u\partial v}=0$.

A complete solution of (\ref{eq:ehj}) consists of the four
combinations of the two independent one-dimensional problems:
\begin{eqnarray}
F(u)&=&\int du \sqrt{2 f(u)} \qquad , \qquad  G(v)=\int dv
\sqrt{2g(v)}\nonumber \\ \xi^*W^{(a,b)}&=&(-1)^a\left( \int du
\sqrt{2 f(u)} +(-1)^b \int dv \sqrt{2g(v)}\right) \quad ,\,
a,b=0,1 \quad .\label{eq:sup1}
\end{eqnarray}
The image of (\ref{eq:ehj})-(\ref{eq:sup1}) in the Cartesian
plane,
\[
U(x^1,x^2)=\frac{1}{2}\frac{\partial W^{(a,b)}}{\partial
x^j}\frac{\partial W^{(a,b)}}{\partial x^j} \quad , \forall
a,b=0,1 \quad , \label{eq:potsup1}
\]
shows that there are four different superpotentials for the same
Type I natural Lagrangian. In Cartesian coordinates the
separability condition $\frac{\partial^2 \xi^*W}{\partial
u\partial v}^{(a,b)}=0$ reads:
\begin{equation}
x^1 x^2 \left( \frac{\partial^2 W^{(a,b)}}{\partial x^1
\partial x^1} -\frac{\partial^2 W^{(a,b)}}{\partial x^2 \partial
x^2}\right)+ (x^2 x^2-x^1 x^1+c^2) \frac{\partial^2
W^{(a,b)}}{\partial x^1 \partial x^2} +x^2 \frac{\partial
W^{(a,b)}}{\partial x^1}-x^1 \frac{\partial W^{(a,b)}}{\partial
x^2}=0 \label{eq:supt1}
\end{equation}

The Hamiltonian
\[
I_1^{(B)}=\frac{1}{2}p_jp_j+\frac{1}{2}\frac{\partial
W^{(a,b)}}{\partial x^j}\frac{\partial W^{(a,b)}}{\partial
x^j}\qquad ,
\]
independent of $a,b$, is an obvious integral of motion. In
elliptic coordinates it reads:
\[
\xi^*I_1^{(B)}={1\over
2(u^2-v^2)}\left\{(u^2-c^2)(p_u^2+2f(u))+(c^2-v^2)(p_v^2+2g(v))\right\}
\qquad ,
\]
where
\[
p_u={u^2-v^2\over u^2-c^2}\dot{u} \qquad , \qquad
p_v={u^2-v^2\over c^2- v^2}\dot{v} \qquad .
\]
The inverse image of the second invariant is:
\begin{eqnarray*}
\xi^* I_2^{(B)} &=& \frac{1}{2} \left[ \frac{(u^2-c^2)(c^2 -
v^2)}{u^2-v^2} p_v^2 -
\frac{(u^2-c^2)(c^2 - v^2) }{u^2 -v^2} p_u^2 \right]\\
&+& \frac{(u^2 -c^2)(c^2 - v^2)}{2 (u^2-v^2)}
\left[\left(\frac{dG}{dv}\right)^2 -
\left(\frac{dF}{du}\right)^2\right]\qquad .
\end{eqnarray*}
The direct image provides the intricate second invariant in
involution - $\{I_1^{(B)},I_2^{(B)}\}_P=0$ - with the Hamiltonian:
\begin{equation}
I_2^{(B)} =\frac{1}{2} \left[ \left(x^2 p_1- x^1 p_2 \right)^2-
c^2 p_2 p_2 + \left(x^2 \frac{\partial W^{(a,b)}}{\partial x^1} -
x^1 \frac{\partial W^{(a,b)}}{\partial x^2}\right)^2- c^2
\frac{\partial W^{(a,b)}}{\partial x^2}\frac{\partial
W^{(a,b)}}{\partial x^2} \right]\quad . \label{eq:seci}
\end{equation}
The remarkable fact is that $I_2^{(B)}$ is independent of $a,b$,
i.e. the four superpotentials also leads to identical second
invariants !, not only to the same Hamiltonian.

In the literature about integrable dynamical systems it is usually
stated that the existence of the second invariant obeys a hidden
symmetry. The definition of the generalized momenta,
\[
\Pi_j^{(a,b)}=p_j+i\frac{\partial W^{(a,b)}}{\partial x^j} \qquad
,
\]
sheds light on the nature of such (non-linear) symmetries. In
terms of the generalized momenta we obtain:
\[
I_1^{(B)}=\frac{1}{2}\left[|\Pi_1^{(a,b)}|^2+
|\Pi_2^{(a,b)}|^2\right]\quad , \quad I_2^{(B)}=\frac{1}{2} \left[
\left| x^2\Pi_1^{(a,b)}-x^1\Pi^{(a,b)}_2 \right|^2-c^2 \left|
\Pi_2^{(a,b)} \right|^2 \right]\quad , \forall a,b \quad ,
\]
and well-known invariants with $p_j$ replaced by $\Pi_j^{(a,b)}$
are recognized. $\Pi_j=\frac{\partial (L+L_T)}{\partial
\dot{x}^j}$ can be understood as the canonical momenta coming from
the addition of a complex topological piece
\[
L_T=i\dot{x}^j\frac{\partial W}{\partial x^j}
\]
to the Lagrangian $L$, in agreement with the central extension
shown in the SUSY Poisson algebra (\ref{eq:exsus}). Note that
$I_1^{(B)}$ can thus be written \`a la Bogomolny:
\[
I_1^{(B)}=\frac{1}{2}\Pi_j\Pi_j-i\Pi_j\frac{\partial W}{\partial
x^j}\qquad .
\]

It is also possible, however, to interpret that the generalized
momenta by themselves close another extension, now of the ordinary
Poisson algebra:
\begin{equation}
\{ x^j,x^k \}_P=0 \qquad , \qquad \{x^j,\Pi_k\}_P=\delta^j_k
\qquad , \qquad \{\Pi_j,\Pi_k\}_P=-2i\frac{\partial ^2 W}{\partial
x^j\partial x^k}\label{eq:poist} \qquad .
\end{equation}
Both the first and the second invariants obey symmetries related
to (non-linear) transformations generated in the framework of the
generalized Poisson structure (\ref{eq:poist}).

\item {\bf Liouville Models of Type II:}

The type II Liouville models are two-dimensional dynamical systems
for which the Hamilton-Jacobi equation is separable using polar
coordinates. The direct - $\zeta :{\Bbb R}^2\longrightarrow {\Bbb
D}^2\simeq {\Bbb R}^2-\{0\}$ - and inverse - $\zeta^{-1} :{\Bbb
R}^2-\{0\}\longrightarrow {\Bbb R}^2$ - maps determine the change
from polar to Cartesian coordinates and viceversa:
\begin{eqnarray*}
(\rho,\chi)&=&\zeta (x^1,x^2)=\left( \sqrt{x^1x^1+x^2x^2},{\rm
arctan}\left(\frac{x^2}{x^1}\right)\right) \\
(x^1,x^2)&=&\zeta^{-1}(\rho,\chi)=\left( \rho \cos \chi ,\rho \sin
\chi\right) \qquad , \, \rho \in [0,\infty); \, \chi\in [0,2\pi)
\quad .
\end{eqnarray*}

The Lagrangian of the Liouville models of Type II reads:
\begin{equation}
\zeta^* L = \displaystyle\frac{1}{2} \, \dot{\rho} \, \dot{\rho} +
\displaystyle\frac{1}{2} \,\rho^2 \,\dot{\chi} \, \dot{\chi}
-f(\rho)-\frac{1}{\rho^2} \, g(\chi) \qquad .\label{eq:liou2}
\end{equation}
Again, besides the metric factor  $g_{11}=1$, $g_{22}=\rho^2$, the
contributions of $\rho$ and $\chi$ appear completely separated in
the Lagrangian. Here, the zero energy time-independent
Hamilton-Jacobi equation (assuming separability) (\ref{eq:spor})
is:
\[
\zeta^* U= f(\rho)+\frac{1}{\rho^2} \, g(\chi) = \frac{1}{2}
\left( \frac{dF}{d\rho} \right)^2\pm \frac{1}{2 \rho^2} \left(
\frac{dG}{d\chi} \right)^2 \qquad .
\]
The complete solution
\begin{eqnarray*}
F(\rho)&=&\int du \sqrt{2 f(\rho)} \qquad , \qquad  G(\chi)=\int
dv \sqrt{2g(\chi)}\nonumber \\ \zeta^*
W^{(a,b)}&=&(-1)^a\left(\int d\rho \sqrt{2 f (\rho)} +(-1)^b \int
d\chi \sqrt{2g (\chi)}\right) \qquad a,b=0,1 \qquad ,
\end{eqnarray*}
provides us with four different superpotentials to build
supersymmetric extensions:
\[
U(x^1,x^2)=\frac{1}{2}\frac{\partial W^{(a,b)}}{\partial
x^j}\frac{\partial W^{(a,b)}}{\partial x^j}\quad , \forall
a,b=0,1 \quad . \label{eq:potsup11}
\]
In Cartesian coordinates the separability condition
$\frac{\partial^2 \zeta^*W}{\partial\rho\partial\chi}^{(a,b)}=0$
reads
\begin{equation}
-x^1 x^2 \left( \frac{\partial^2 W^{(a,b)}}{\partial x^1
\partial x^1}- \frac{\partial^2 W^{(a,b)}}{\partial x^2 \partial
x^2}\right)+(x^1 x^1-x^2 x^2) \frac{\partial^2 W^{(a,b)}}{\partial
x^1
\partial x^2}+x^1 \frac{\partial W^{(a,b)}}{\partial x^2}-x^2
\frac{\partial W^{(a,b)}}{\partial x^1}=0 \label{eq:supt2}
\end{equation}
The two invariants in involution, written in polar coordinates,
are:
\[
\zeta^*I_1^{(B)}=\frac{1}{2}p_\rho^2+\frac{1}{2\rho^2}p_\chi^2+f(\rho)+\frac{1}{\rho^2}g(\chi)
\quad , \quad
\zeta^*I_2^{(B)}=\frac{1}{2\rho^2}p_\chi^2+\frac{1}{\rho^2}g(\chi)
\quad .
\]
In Cartesian coordinates they are easily shown to be independent
of $a$ and $b$:
\begin{eqnarray*}
I_1^{(B)}&=&\frac{1}{2}p_jp_j+\frac{1}{2}\frac{\partial
W^{(a,b)}}{\partial x^j}\frac{\partial W^{(a,b)}}{\partial
x^j}=\frac{1}{2}|\Pi_j^{(a,b)}||\Pi_j^{(a,b)}|
\\ I_2^{(B)} &=&  \frac{1}{2} \left(x^2 \dot{x}^1 - x^1 \dot{x}^2
\right)^2+ \frac{1}{2} \left(x^2 \frac{\partial
W^{(a,b)}}{\partial x^1} - x^1 \frac{\partial W^{(a,b)}}{\partial
x^2} \right)^2=\frac{1}{2}|x^2\Pi_1^{(a,b)}-x^1\Pi_2^{(a,b)}|^2
\qquad .
\end{eqnarray*}

\item {\bf Liouville Models of Type III :}

In this type of model, the Hamilton-Jacobi equation is separable
using parabolic coordinates. The direct - $\gamma :{\Bbb
R}^2\longrightarrow {\Bbb H}^2$ - and inverse - $\gamma^{-1}
:{\Bbb H}^2\longrightarrow {\Bbb R}^2$ - maps between the
half-plane and the plane dictate the change from parabolic to
Cartesian coordinates and viceversa:
\begin{eqnarray*}
&&(u,v)=\gamma (x^1,x^2)=\left(
\pm\sqrt{\sqrt{x^1x^1+x^2x^2}+x^1},
\sqrt{\sqrt{x^1x^1+x^2x^2}-x^1}\right) \\
&&(x^1,x^2)=\gamma^{-1}(u,v)=\left( \frac{1}{2}(u^2-v^2) , uv\right) \\
&&{\Bbb H}^2=(-\infty,\infty)\times [0,\infty) \qquad ; \qquad
u\in (-\infty,\infty) \, , \, v\in [0,\infty) \qquad .
\end{eqnarray*}

A Liouville model of Type III obeys a Lagrangian of the form:
\begin{equation}
\gamma^* L = \displaystyle\frac{1}{2} (u^2+v^2) \left( \dot{u} \, \dot{u} +\dot{v} \,
\dot{v} \right)
-\displaystyle\frac{1}{u^2+v^2} \left( f(u)+g(v) \right)
\label{eq:liou3}
\end{equation}
The zero-energy static Hamilton-Jacobi equation
\[
\gamma^*U=\frac{1}{u^2+v^2}\left(f(u)+g(v)\right)=
\frac{1}{u^2+v^2}\left[\left(\frac{dF}{du}\right)^2+\left(\frac{dG}{dv}\right)^2\right]
\]
is solved by the four ``separate'' superpotentials:
\[
\gamma^* W^{(a,b)}= (-1)^a\left( \int du\sqrt{2 f(u)} +(-1)^b
\int dv \sqrt{2g(v)}\right)\qquad .
\]
The separability condition $\frac{\partial^2 W}{\partial v\partial
u}^{(a,b)}=0$ in Cartesian coordinates takes the form:
\begin{equation}
x^2 \left( \frac{\partial^2 W^{(a,b)}}{\partial x^2
\partial x^2}- \frac{\partial^2 W^{(a,b)}}{\partial x^1 \partial
x^1}\right)+ 2 x^1 \frac{\partial^2 W^{(a,b)}}{\partial x^1
\partial x^2}+\frac{\partial W^{(a,b)}}{\partial x^2}=0 \label{eq:supt3}
\end{equation}
In parabolic coordinates the two invariants in involution can be
chosen as:
\begin{eqnarray*}
\gamma^*I_1^{(B)}&=&\frac{1}{2(u^2+v^2)}\left(p_u^2+p_v^2\right)+
\frac{1}{u^2+v^2}\left(f(u)+g(v)\right)\\
\gamma^*I_2^{(B)}&=&\frac{1}{u^2+v^2}\left(\frac{u^2}{2}p_v^2-\frac{v^2}{2}p_u^2
+u^2g(v)-v^2f(u)\right)\qquad .
\end{eqnarray*}
They can easily be translated to Cartesian coordinates, giving the
same result for all the values of $a$ and $b$:
\begin{eqnarray*}
I_1^{(B)}&=&\frac{1}{2}p_jp_j+\frac{1}{2}\frac{\partial
W^{(a,b)}}{\partial x^j}\frac{\partial W^{(a,b)}}{\partial
x^j}=\frac{1}{2}|\Pi_j^{(a,b)}||\Pi_j^{(a,b)}|\\
I_2^{(B)}&=&\left(x^1 p_2 - x^2 p_1 \right) p_2+ \left(x^1
\frac{\partial W^{(a,b)}}{\partial x^2} - x^2 \frac{\partial
W^{(a,b)}}{\partial x^1} \right)\frac{\partial W^{(a,b)}}{\partial
x^2}\\&=&{\rm Re}\left[(x^2\Pi_1^{(a,b)}
-x^1\Pi_2^{(a,b)})\bar{\Pi}_2^{(a,b)}\right] \qquad .
\end{eqnarray*}
\item {\bf Liouville Models of Type IV:}

In the fourth type of Liouville models the dynamics of the two
degrees of freedom are completely independent:
\begin{equation}
L = \displaystyle\frac{1}{2} \,\dot{x}^1 \, \dot{x}^1 \,+\,
\displaystyle\frac{1}{2} \, \dot{x}^2 \, \dot{x}^2\,
-\,f(x^1)\,-\,g(x^2) \label{eq:liou4}
\end{equation}
In this case there is no need to change the system of coordinates
to solve the HJ equation and it is clear that the four
superpotentials are:
\[
W^{(a,b)}(x^1,x^2)=(-1)^a\left( \int dx^1 \sqrt{f(x^1)} +(-1)^b
\int dx^2 \sqrt{g(x^2)}\right)\qquad .
\]
The separability condition is
\begin{equation}
\frac{\partial^2 W^{(a,b)}}{\partial x^1 \partial x^2}=0
\label{eq:supt4}
\end{equation}
and the two invariants can be chosen as:
\begin{eqnarray*}
I_1^{(B)}&=&\frac{1}{2}p_jp_j+\frac{1}{2}\frac{\partial
W^{(a,b)}}{\partial x^j}\frac{\partial W^{(a,b)}}{\partial
x^j}=\frac{1}{2}|\Pi_j^{(a,b)}||\Pi_j^{(a,b)}|\\ I_2^{(B)}&=&
\frac{1}{2} p_1 p_1 +\frac{1}{2} \frac{\partial
W^{(a,b)}}{\partial x^1} \frac{\partial W^{(a,b)}}{\partial
x^1}=\frac{1}{2}|\Pi_1^{(a,b)}|.|\Pi_1^{(a,b)}| \qquad .
\end{eqnarray*}
\end{itemize}
\subsection{SuperLiouville models}

The strategy for building ${\cal N}=2$ supersymmetric Lagrangian
systems with Liouville models as the bosonic sector is clear:
simply consider the hybrid non-linear Sigma/Wess-Zumino models of
subsection \S 2.3 whose target space is the Riemannian manifold
${\Bbb M}^2=\bar{\Bbb D}^2$ equipped with the metric induced by
the maps $\xi^*$, $\zeta^*$ and $\gamma^*$ for Types I, II, and
III, and the Euclidean metric for Type IV.

\begin{itemize}
\item {\bf  SuperLiouville Models of Type I:}
The metric and the Christoffel symbols induced by the map $\xi^*$
in ${\Bbb D}^2=(c,\infty)\times (-c,c )$ are:
\[
g(u,v)=\left( \begin{array}{cc}
g_{uu}=\displaystyle\frac{u^2-v^2}{u^2-c^2} & g_{uv}=0 \\
g_{vu}=0 & g_{vv}=\displaystyle\frac{u^2-v^2}{c^2-v^2}\end{array}
\right)
\]
\[
g^{-1}(u,v)=\left( \begin{array}{cc}
g^{uu}=\displaystyle\frac{u^2-c^2}{u^2-v^2} & g^{uv}=0 \\
g^{vu}=0 & g^{vv}=\displaystyle\frac{c^2-v^2}{u^2-v^2}
\end{array} \right)
\]
\[
\begin{array}{lclcl}
\Gamma_{uu}^u=\displaystyle\frac{-u
(c^2-v^2)}{(u^2-v^2)(u^2-c^2)} & ,&
\Gamma_{vv}^v=\displaystyle\frac{v (u^2-c^2)}{(u^2-v^2)(c^2-v^2)}
& ,& \Gamma_{uv}^u=\Gamma_{vu}^u=\displaystyle\frac{-v}{u^2-v^2}
\\[0.4cm] \Gamma_{uu}^v=\displaystyle\frac{v
(c^2-v^2)}{(u^2-v^2)(u^2-c^2)} & ,&
\Gamma_{vv}^u=\displaystyle\frac{-u
(u^2-c^2)}{(u^2-v^2)(c^2-v^2)} &,&
\Gamma_{uv}^v=\Gamma_{vu}^v=\displaystyle\frac{u }{u^2-v^2}
\end{array}
\]
Besides the bosonic (even Grassman) variables $u$ , $v$ , ruled by
Lagrangians of Type I as discussed in the previous sub-Section,
there are also fermionic (odd Grassman) Majorana spinors
$\vartheta^u_\alpha$ , $\vartheta^v_\alpha$ in the system. A
supersymmetric two dimensional mechanical system is a super-
Liouville model of Type I if the Lagrangian is of the form
$\xi^*L=\xi^*L_B+\xi^*L_F+\xi^*L_{BF}$ with:
\[
\xi^*L_B={1\over 2}g_{uu}(u,v)\dot{u}\dot{u}+{1\over
2}g_{vv}(u,v)\dot{v}\dot{v}-{1\over 2}g^{uu}(u,v)\left({dF\over
du}\right)^2-{1\over 2}g^{vv}(u,v)\left({dG\over dv}\right)^2
\qquad ,
\]
\[
\xi^*L_F=\frac{i}{2} g_{uu}(u,v) \vartheta_\alpha^u D_t
\vartheta_\alpha^u+ \frac{i}{2} g_{vv}(u,v)\vartheta_\alpha^v D_t
\vartheta_\alpha^v \qquad ,
\]
and,
\begin{eqnarray*}
\xi^*L_{BF}^I & =& -i \left[ \frac{d^2 F}{du^2}-\Gamma^u_{uu}
\frac{dF}{du} -\Gamma^v_{uu}\frac{dG}{dv} \right] \vartheta_1^u
\vartheta_2^u  -i \left[ \frac{d^2 G}{dv^2} - \Gamma^u_{vv}
\frac{dF}{du} -\Gamma^v_{vv} \frac{dG}{dv}  \right] \vartheta_1^v
\vartheta_2^v+
\\ & & +i \left[ \Gamma^u_{uv}
\frac{dF}{du}-\Gamma^v_{uv} \frac{dG}{dv}  \right] (\vartheta_1^u
\vartheta_2^v+\vartheta_1^v \vartheta_2^u) \qquad .
\end{eqnarray*}
The fermionic kinetic energy is encoded in $\xi^*L_F$ and there are
Yukawa terms in $\xi^*L_{BF}$ ruling the Bose-Fermi interactions.

\item {\bf SuperLiouville Models of Type II:}
Everything is analogous trading elliptic for polar coordinates.
The metric and the Christoffel symbols are:
\[
g(\rho)=\left(
\begin{array}{cc} g_{\rho\rho}=1 & g_{\rho\chi}=0 \\ g_{\chi\rho}=0 & g_{\chi\chi}=\rho^2
\end{array}  \right)\quad , \quad g^{-1}(\rho)=\left(
\begin{array}{cc} g^{\rho\rho}=1 & g^{\rho\chi}=0 \\ g^{\chi\rho}=0 &
g^{\chi\chi}={1\over\rho^2}
\end{array}  \right)
\]
\[
\Gamma_{\rho\rho}^\rho=\Gamma_{\rho\chi}^\rho=\Gamma_{\chi\rho}^\rho
=\Gamma_{\rho\rho}^\chi=\Gamma_{\chi\chi}^\chi= 0\  ,\hspace{1cm}
\Gamma_{\chi\chi}^\rho=-\rho \ ,\hspace{1cm}
\Gamma_{\rho\chi}^\chi=\Gamma_{\chi\rho}^\chi=\frac{1}{\rho}
\]
Therefore, the Lagrangian of a super-Liouville model of Type II
including bosonic $\rho$ , $\chi$ and fermionic
$\vartheta^\rho_\alpha$ , $\vartheta^\chi_\alpha$ variables is the
sum of the three pieces:
\[
\zeta^*L_B= \displaystyle\frac{1}{2} \dot{\rho} \dot{\rho} +
\frac{1}{2} g_{\chi\chi} \dot\chi \dot\chi -
\displaystyle\frac{1}{2} \left( \displaystyle\frac{dF}{d\rho}
\right)^2 -\frac{1}{2}g^{\chi\chi} \left(
\displaystyle\frac{dG}{d\chi} \right)^2
\]
\[
\zeta^*L_F=\frac{i}{2} \vartheta_\alpha^{\rho} D_t
\vartheta_\alpha^{\rho}+ \frac{i}{2} g_{\chi\chi}
\vartheta_\alpha^\chi D_t \vartheta_\alpha^\chi
\]
\[
\zeta^*L_{BF}= -i \frac{d^2 F}{d \rho^2} \vartheta_1^\rho
\vartheta_2^\rho- i \left( \frac{d^2 G}{d\chi^2}
-\Gamma^\rho_{\chi\chi} \frac{dF}{d\rho} \right)
\vartheta_1^\chi\vartheta_2^\chi +i \Gamma^\chi_{\rho\chi}
\frac{dG}{d\chi} (\vartheta_1^\rho
\vartheta_2^\chi+\vartheta_1^\chi \vartheta_2^\rho ) \qquad .
\]

\item {\bf SuperLiouville Models of Type III:}
The supersymmetric extensions of Liouville Type III models is
easier in parabollic coordinates. The metric and the Christoffel
symbols are:
\[
g(u,v)=\left(
\begin{array}{cc} u^2+v^2 & 0 \\ 0 & u^2+v^2 \end{array}  \right)
\qquad , \qquad g^{-1}(u,v)=\left(
\begin{array}{cc} {1\over u^2+v^2} & 0 \\ 0 & {1\over u^2+v^2} \end{array}\right)
\]
\[
\Gamma_{uu}^u=\Gamma_{uv}^v=\Gamma_{vu}^v=-\Gamma_{vv}^u=
\displaystyle\frac{u}{u^2+v^2} \  ,\hspace{1cm}
\Gamma_{uv}^u=\Gamma_{vu}^u=\Gamma_{vv}^v=-\Gamma_{uu}^v=
\displaystyle\frac{v}{u^2+v^2}
\]
The dynamics of the SUSY pairs of variables $u$ ,
$\vartheta^u_\alpha$ and $v$ , $\vartheta^v_\alpha$ is governed in
a super-Liouville model of Type III by the Lagrangian
$\zeta^*L=\zeta^*L_B+\zeta^*L_F+\zeta^*L_{BF}$, where
\[
\zeta^*L_B= \displaystyle\frac{1}{2} g_{uu}\dot{u} \dot{u}+
{1\over 2}g_{vv}\dot{v} \dot{v}  -
\displaystyle\frac{1}{2}g^{uu}\left( \displaystyle\frac{dF}{du}
\right)^2+ {1\over 2}g^{vv}\left( \displaystyle\frac{dG}{dv}
\right)^2 \qquad ,
\]
\[
\zeta^*L_F=\frac{i}{2} g_{uu} \vartheta_\alpha^u D_t
\vartheta_\alpha^u+ \frac{i}{2} g_{vv} \vartheta_\alpha^v D_t
\vartheta_\alpha^v \qquad ,
\]
and,
\begin{eqnarray*}
\zeta^*{L}_{BF}&=& -i \left[ \frac{d^2F}{du^2}
-\Gamma^u_{uu}\frac{dF}{du} -\Gamma^v_{uu} \frac{dG}{dv}  \right]
\vartheta_1^u \vartheta_2^u -i \left[ \frac{d^2 G}{dv^2}
-\Gamma^u_{vv}\frac{dF}{du} - \Gamma^v_{vv}\frac{dG}{dv}  \right]
\vartheta_1^v \vartheta_2^v +
\\ & & +i \left[ \Gamma^v_{uv}
\frac{dG}{dv} + \Gamma^u_{uv}\frac{dF}{du}  \right] (\vartheta_1^u
\vartheta_2^v+\vartheta_1^v \vartheta_2^u) \qquad .
\end{eqnarray*}

\noindent $\bullet$ {\bf SuperLiouville Models of Type IV:}
Finally, the definition of SuperLiouville Model of Type IV is
straightforward. The Lagrangian is
\[
L = \displaystyle\frac{1}{2} \dot{x}^j \dot{x}^j +
\displaystyle\frac{i}{2} \theta_\alpha^j \dot{\theta}_\alpha^j
-\displaystyle\frac{1}{2} \displaystyle\frac{\partial W}{\partial x^j}
\displaystyle\frac{\partial W}{\partial x^j}
-i \displaystyle\frac{\partial^2 W}{\partial x^1 \partial x^1} \theta_1^1 \theta_2^1
-i \displaystyle\frac{\partial^2 W}{\partial x^2 \partial x^2} \theta_1^2 \theta_2^2
\]
and the system can be understood as an $\left( {\cal N}=2\right)
\oplus \left( {\cal N}=2\right)$ SUSY model in $(0+1)$ dimensions.

\vspace*{0.1cm}
\end{itemize}

There is an obvious first integral that can be written in a
unified way for all four Types of superLiouville models using
Cartesian coordinates:
\[
I_1=I_1^{(B)}+I_1
^{(F)}\qquad ,\qquad
I_1^{(B)}=\frac{1}{2}p_jp_j+\frac{1}{2}\frac{\partial
W^{(a,b)}}{\partial x^j}\frac{\partial W^{(a,b)}}{\partial x^j}
\qquad ,  \qquad I_1^{(F)}=i\frac{\partial ^2 W^{(a,b)}}{\partial
x^j\partial x^k}\theta^j_1\theta^k_2 \qquad .
\]
$I_1^{(B)}$ is formally identical to the Hamiltonian of the parent
Liouville model but we stress that $p_j$ and $x^j$ are now even
Grassman variables. It is, in any case, independent of $a$ and
$b$. $I_1^{(F)}$ comes from the Yukawa couplings between bosonic
and fermionic variables. Note that $I_1^{(F)}$ depends on choosing
either $b=0$ or $b=1$; thus, each Liouville model admits two
different supersymmetric extensions achieved from different
solutions of the time-independent Hamilton-Jacobi equation for the
Hamilton characteristic function (the superpotential). The choice
of $a=1$ instead of $a=0$ changes $I_1^{(F)}$ to $-I_1^{(F)}$;
this flip of sign can be absorbed by exchanging positive and
negative energy for the fermionic trajectories.

A final remark is that the separability of the purely bosonic
Liouville mechanical systems is lost in the supersymmetric
framework because of the Yukawa couplings, except for
SuperLiouville models of Type IV.

\section{On the Bosonic Invariants}
Liouville models have a second invariant in involution with the
energy -the first invariant- that guarantees complete
integrability in the sense of the Liouville theorem. We shall now
show that the SuperLiouville models also have a second invariant
of bosonic nature. Our strategy in the search for such an
invariant, $\{I,H\}_P=0$, follows the general pattern found in the
literature: see \cite{Hietarinta}. The ansatz for invariants, at
most quadratic in the momenta, is:
\[ \begin{array}{lcr}
I&=&\displaystyle{\frac{1}{2} H^{ij} p_i p_j + K(x^1,x^2)+F_{ij}
\theta_1^i \theta_2^j+G_{ij} \theta_1^i \theta_1^j+ J_{ij}
\theta_2^i \theta_2^j+} \\ & & \displaystyle{+ L^i_{jk} p_i
\theta_1^j \theta_1^k+ M^i_{jk} p_i \theta_2^j \theta_2^k+
N^i_{jk} p_i \theta_1^j \theta_2^k+ S_{ijkl} \theta_1^i \theta_2^k
\theta_1^j \theta_2^l}
\end{array}
\]
Here, we assume that:
\begin{enumerate}
\item[$i)$] $K$ is a function.

\item[$ii)$] $H^{ij}$ is a symmetric tensor depending on $x^i$. There are
three independent functions to determine.

\item[$iii)$] $L^i_{jk}$ and $M^i_{jk}$ also depends only on $x^i$ and are
antisymmetric in the indices $j$ and $k$: $L^i_{jk}=-L^i_{kj}$,
$M^i_{jk}=-M^i_{kj}$. They include four independent functions.

\item[$iv)$] $G_{ij}$ and $J_{ij}$ are functions, antisymmetric in the
indices, of $x^i$: $G_{ij}=-G_{ji}$ and $J_{ij}=-J_{ji}$. A priori
$F_{ij}(x^i)$ however, is neither symmetric nor antisymmetric; it
contains four independent functions.

\item[$v)$] Finally, $S_{ijkl}(x^i)$ is antisymmetric in the exchange of the
indices $i,j$ and $k,l$ and symmetric in the exchange of the pairs
$ij,kl$. There is only one independent function to determine in
this tensor.

\end{enumerate}

\noindent The commutator with the Hamiltonian is:
\begin{eqnarray*}
 [I,H]& =& {\displaystyle - \frac{1}{2} \frac{\partial H^{jk}}{\partial x^l} p_l
p_j p_k +\left( H^{lj} \frac{\partial^2 W}{\partial x^j\partial
x^k} \frac{\partial W}{\partial x^k}-\frac{\partial K}{\partial
x^l} \right) p_l +}
\\
& & \hspace{-0.6cm} {\displaystyle + \left(i H^{nj}
\frac{\partial^3 W}{\partial x^k\partial x^l\partial x^j}-
\frac{\partial F_{kl}}{\partial x^n}-2 L_{km}^n\frac{\partial^2
W}{\partial x^m\partial x_l}-2  M_{lm}^n \frac{\partial^2
W}{\partial x^m\partial x^k} \right) p_n
\theta_1^k \theta_2^l+} \\
& & \hspace{-0.6cm} {\displaystyle  + \left(-\frac{\partial^2
W}{\partial x^l \partial x^k} F_{lj}+ M^n_{kj} \frac{\partial^2
W}{\partial x^n\partial x^l}\frac{\partial W}{\partial x^l}
\right) \theta_2^k \theta_2^j +\left(- \frac{\partial
J_{lj}}{\partial x^k}-N^k_{mj} \frac{\partial^2 W}{\partial
x^m\partial x^l}
\right) p_k \theta_2^l \theta_2^j-}\\
& & \hspace{-0.6cm} {\displaystyle + \left(\frac{\partial^2
W}{\partial x^j\partial x^k} F_{nk}+ L^l_{nj} \frac{\partial^2 W
}{\partial x^l\partial x^k}\frac{\partial W}{\partial x^k} \right)
\theta_1^n \theta_1^j+ \left( -\frac{\partial G_{nj}}{\partial
x^l}+N^l_{nk}\frac{\partial^2 W}{\partial x^j\partial x^k} \right)
p_l \theta_1^n \theta_1^j+}
\\
& & \hspace{-0.6cm} {\displaystyle  +\, 2 \left(- G_{nj}
\frac{\partial^2 W}{\partial x^j\partial x^k}- J_{kl}
\frac{\partial^2 W}{\partial x^n\partial x^l}+\frac{1}{2} N^j_{nk}
\frac{\partial^2 W}{\partial x^j\partial x^l}\frac{\partial
W}{\partial x^l} \right) \theta_1^n \theta_2^k+}
\\
& & \hspace{-0.6cm} {\displaystyle - \frac{\partial
L^n_{jk}}{\partial x^l} p_l p_n \theta_1^j \theta_1^k-
\frac{\partial M^n_{jk}}{\partial x^l} p_l p_n \theta_2^j
\theta_2^k- \frac{\partial N^n_{jk}}{\partial x^l}
p_l p_n \theta_1^j \theta_2^k+} \\
& & \hspace{-0.6cm} {\displaystyle +i N^n_{jk} \frac{\partial^3
W}{\partial x^n\partial x^l\partial x^m} \theta_1^j \theta_2^k
\theta_1^l \theta_2^m -\frac{\partial S_{njkl}}{\partial x^m} p_m
\theta_1^n \theta_2^k \theta_1^j \theta_2^l }
\end{eqnarray*}
\noindent The relationship $\{I,H\}_P=0$ guarantees that $I$ will
be an invariant of the supersymmetric mechanical system.
Therefore, conditions making the Poisson bracket vanish (shown in
the table 1) must be imposed.

\begin{table}[htbp]
\begin{center}
\begin{tabular}{|l|l|}
\hline &
\\
{\small BOX 1} & \mbox{\bf a)}
$\displaystyle \sum_{[ijk]} \frac{\partial H^{ij}}{\partial x^k}=0$ \\ & \\
\hline & \\ {\small BOX 2} & \mbox{\bf a)} $\displaystyle H^{ij}
\frac{\partial^2 W}{\partial x^j\partial x^k}\frac{\partial W}{\partial x^k}=
\frac{\partial K}{\partial x^i}$  \\ & \\
\hline & \\  {\small BOX 3} & \mbox{\bf a)}
$\displaystyle{\epsilon^{jk} \frac{\partial L^i_{jk}}{\partial
x^l}+\epsilon^{jk} \frac{\partial L^l_{jk}}{\partial x^i} =0 }$ \\
& \mbox{\bf b)} $ \displaystyle{\epsilon^{jk} \frac{\partial
M^i_{jk}}{\partial x^l}+\epsilon^{jk}
\frac{\partial M^l_{jk}}{\partial x^i} =0}$   \\ & \\
\hline &\\  {\small BOX 4} & \mbox{\bf a)}
$\displaystyle{H^{nj}\frac{\partial^3 W}{\partial x^k\partial
x^l\partial x^j}+i \frac{\partial F_{kl}}{\partial x^n}+2 i
L_{km}^n\frac{\partial^2 W}{\partial x^m\partial x^l}+2 i M_{lj}^n
\frac{\partial^2 W}{\partial x^j\partial x^k}=0}$ \\ & \mbox{\bf
b)} $\displaystyle{\epsilon^{ij}\left( \frac{\partial^2
W}{\partial x^i \partial x^k} F_{kj}- M^k_{ij} \frac{\partial^2
W}{\partial x^k \partial x^l}\frac{\partial W}{\partial x^l}
\right) =0}$ \\ & \mbox{\bf c)} $\displaystyle{\epsilon^{ij}
\left(\frac{\partial^2 W}{\partial x^j \partial x^k} F_{ik}+
L^l_{ij} \frac{\partial^2 W}{\partial x^k\partial
x^l}\frac{\partial
W}{\partial x^k} \right)=0}$ \\ & \\ \hline & \\
{\small BOX 5}  & \mbox{\bf a)} $\displaystyle \frac{\partial
N^i_{jk}}{\partial x^l}+ \frac{\partial N^l_{jk}}{\partial x^i}=0
$\\ & \mbox{\bf b)} $\displaystyle \epsilon^{ij} \epsilon^{lk}
N^m_{jk}\frac{\partial^3 W}{\partial x^i\partial x^l\partial
x^m}=0 $ \\ &\\ \hline &\\
{\small BOX 6} & \mbox{\bf a)} $\displaystyle\epsilon^{ij} \left(
\frac{\partial G_{ij}}{\partial x^l}-N^l_{ik}\frac{\partial^2
W}{\partial x^j \partial x^k} \right)=0$
\\ &\mbox{\bf b)} $\displaystyle \epsilon^{ij} \left( \frac{\partial J_{ij}}{\partial
x^k}+N^k_{mj} \frac{\partial^2 W}{\partial x^m\partial x^i}
\right)=0 $ \\ &\mbox{\bf c)} $\displaystyle
G_{ij}\frac{\partial^2 W}{\partial x^j\partial x^k}+
J_{kl}\frac{\partial^2 W}{\partial x^i\partial x^l}-\frac{1}{2}
N^j_{ik} \frac{\partial^2 W}{\partial x^j\partial
x^l}\frac{\partial W}{\partial x^l} =0 $
\\  &\\ \hline & \\
{\small BOX 7} & \mbox{\bf a)} $\displaystyle \epsilon^{ij}
\epsilon^{kl} \frac{\partial S_ {ijkl}}{\partial x^m}=0$ \\ & \\
\hline
\end{tabular}
\end{center}
\caption{{\it Conditions on $I$ to obtain a supersymmetric invariant.}}
\end{table}

The sum in expression a) in the box 1 ranges over all the
permutations of the indices $i,j$ and $k$. We deal with a
overdetermined system of partial differential equations: there are
31 PDE relating 15 unknown functions. In the table 1, we have
organized the conditions in a step-by-step distribution, i.e. ,
generically solving conditions in a given box is required to solve
the conditions in the following box. In boxes 1 and 2, the
equations handled by Hietarinta in the bosonic sector are
reproduced. Note too the possibility of the existence of several
solutions, giving rise to different supersymmetric invariants.

\subsection{Second invariant in SuperLiouville models}

In order to solve the above conditions, we proceed in a recurrent
way:

\begin{enumerate}
\item[$i)$] The equations in {\it BOXES 1} and {2} are sufficient to find
$H^{ij}$ and $K$. We recover the information about the second
invariant of the purely bosonic sector.

\item[$ii)$] The equations in {\it BOX 3} are solved if the independent components
of $L^i_{jk}$ and $M^i_{jk}$ are of the form,
\[
L^i_{12}=C\, \epsilon^{ij} x^j +A_i \hspace{2cm} M^i_{12}=D \,
\epsilon^{ij} x^j +B_i
\]
where $A_i$, $B_i$, $C$ y $D$ are constants.

\item[$iii)$] The equations in {\it BOX 4}, together with the previous
information, lead to the computation of $F_{ij}$. The existence of
a solution in equation 4a) requires compliance with the identity
$\frac{\partial^2 F_{kl}}{\partial x^1\partial
x^2}=\frac{\partial^2 F_{kl}}{\partial x^2\partial x^1}$, which
becomes
\begin{equation}
\epsilon^{mn} \displaystyle\frac{\partial}{\partial x^m} \left[
L^n_{jk} \frac{\partial^2 W}{\partial x^j\partial x^l}+
M^n_{jl}\frac{\partial^2 W}{\partial x^k\partial x^j}+\frac{i}{2}
H^{nj}\frac{\partial^3 W}{\partial x^j\partial x^k\partial x^l}
\right]=0 \label{eq:4d}
\end{equation}
Moreover, if we restrict $F_{ij}$ to be symmetric under the
exchange of indices and identify $L^i_{jk}=M^i_{jk}$, equation 4b)
becomes equal to 4c). Keeping in mind the formulae
(\ref{eq:supt1}), (\ref{eq:supt2}), (\ref{eq:supt3}) and
(\ref{eq:supt4}), this latter condition and (\ref{eq:4d}) become
identities, choosing in each type of Liouville models the
appropriate values of the integration constants $A_i$, $B_i$, $C$
and $D$, obtained in the previous point. Thus, the compatibility
of the equations is satisfied and we obtain finally $F_{ij}$ from
formula a).

\item[$iv)$] The equations of {\it BOXES 5, 6} and {\it 7} are satisfied if we consider:
\[
G_{ij}=J_{ij}=N_{ijk}=S_{ijkl}=0
\]

\end{enumerate}

The second invariant in superLiouville models is of the form
\[
I_2=I_2^{(B)}+I_2^{(F)}
\]
where $I_2^{(B)}$ agrees with the purely bosonic second invariant
but bearing in mind that the variables have an even Grassmannian
character and $I_2^{(F)}$ includes terms also involving the
fermionic variables $\theta_\alpha^i$. We find:

\subsubsection{SuperLiouville Models of Type I:}
\begin{eqnarray}
I_2^{(B)}& =&\frac{1}{2} \left[ \left(x^2 p_1- x^1 p_2 \right)^2-
c^2 p_2 p_2 + \left(x^2 \frac{\partial W^{(a,b)}}{\partial x^1} -
x^1 \frac{\partial W^{(a,b)}}{\partial x^2}\right)^2- c^2
\frac{\partial W^{(a,b)}}{\partial x^2}\frac{\partial
W^{(a,b)}}{\partial x^2} \right] \nonumber\\ I_2^{(F)}&=&i (x^2
p_1 - x^1 p_2) \theta_\alpha^1 \theta_\alpha^2+i\left(2 x^1
\frac{\partial^2 W^{(a,b)}}{\partial x^2
\partial x^2}-\frac{\partial W^{(a,b)}}{\partial x^1}-x^2 \frac{\partial^2
W^{(a,b)}}{\partial x^1 \partial x^2} \right) \theta_1^2
\theta_2^2 +\nonumber
\\ & +&i \, \left(-x^1 x^2 \frac{\partial^2 W^{(a,b)}}{\partial x^2
\partial x^2}+x^2 \frac{\partial W^{(a,b)}}{\partial x^1}+ x^2 x^2
\frac{\partial^2 W^{(a,b)}}{\partial x^1 \partial x^2}\right)
(\theta_1^1 \theta_2^2+\theta_1^2 \theta_2^1) +\\ &  +&i \,
\left(- x^2 \frac{\partial W^{(a,b)}}{\partial x^2}- x^1 x^2
\frac{\partial^2 W^{(a,b)}}{\partial x^1 \partial x^2}+ x^2 x^2
\frac{\partial^2 W^{(a,b)}}{\partial x^1 \partial x^1} \right)
\theta_1^1 \theta_2^1 \label{eq:sinvI}
\end{eqnarray}

\subsubsection{SuperLiouville Models of Type II:}
\begin{eqnarray}
I_2^{(B)}& = & \frac{1}{2} \left(x^2 p_1 - x^1 p_2 \right)^2+
\frac{1}{2} \left(x^2 \frac{\partial W^{(a,b)}}{\partial x^1} -
x^1 \frac{\partial W^{(a,b)}}{\partial x^2} \right)^2\nonumber
\\
I_2^{(F)}&=& i \left( x^2 p_1 - x^1 p_2 \right) \theta_\alpha^1
\theta_\alpha^2+i x^2 \left(x^2 \frac{\partial W^{(a,b)}}{\partial
x^1
\partial x^1}-x^1 \frac{\partial^2 W^{(a,b)}}{\partial x^1
\partial x^2}-\frac{\partial W^{(a,b)}}{\partial x^2}\right) \theta_1^1
\theta_2^1+\nonumber \\ & +& i x^2\left(x^2 \frac{\partial^2
W^{(a,b)}}{\partial x^1 \partial x^1}+\frac{\partial
W^{(a,b)}}{\partial x^1}-x^1 \frac{\partial^2 W^{(a,b)}}{\partial
x^2
\partial x^2}\right)(\theta_1^1 \theta_2^2+\theta_1^2
\theta_2^1)+\nonumber
\\ &+& i x^1 \left(x^1 \frac{\partial^2 W^{(a,b)}}{\partial x^2  \partial
x^2}-\frac{\partial W^{(a,b)}}{\partial x^1}-x^2 \frac{\partial^2
W^{(a,b)}}{\partial x^1 \partial x^2}\right) \theta_1^2 \theta_2^2
\label{eq:sinvII}
\end{eqnarray}

\subsubsection{SuperLiouville Models of Type III:}
\begin{eqnarray}
I_2^{(B)}&=&\left(x^1 p_2 - x^2 p_1 \right) p_2+ \left(x^1
\frac{\partial W^{(a,b)}}{\partial x^2} - x^2 \frac{\partial
W^{a,b)}}{\partial x^1} \right)\frac{\partial W^{(a,b)}}{\partial
x^2} \nonumber\\ I_2^{(F)}&=&- i p_2 \theta_\alpha^1
\theta_\alpha^2- i x^2 \frac{\partial^2 W^{(a,b)}}{\partial x^1
\partial x^2} \theta_1^1 \theta_2^1-i x^2 \frac{\partial^2
W^{(a,b)}}{\partial x^2
\partial x^2} (\theta_1^1 \theta_2^2+\theta_1^2 \theta_2^1) +\nonumber\\ &
&+i\left(2 x^1 \frac{\partial^2 W^{(a,b)}}{\partial x^2 \partial
x^2}-\frac{\partial W^{(a,b)}}{\partial x^1}-x^2 \frac{\partial^2
W^{(a,b)}}{\partial x^1 \partial x^2}\right) \theta_1^2 \theta_2^2
\label{eq:sinvIII}
\end{eqnarray}

We observe a common feature in the second invariant of
superLiouville models of Type I, II, and III:
$l^{12}=x^1p_2-x^2p_1$ can be replaced by
$j^{12}=x^1p_2-x^2p_1+i\theta_\alpha^1\theta_\alpha^2$ - invariant
only if there is rotational symmetry - and, thus,
$s^{12}=i\theta^1_\alpha\theta^2_\alpha$ can be interpreted as the
spin of the supersymmetric particle. By adding $S_3$ to the second
invariant in models of Type I and II we obtain a new invariant in
the form of:
\[
I_2^\prime=I_2^{(B)}+I_2^{(F)}+\frac{1}{4}S_3 \qquad .
\]
There is a term,
\[
{1\over 2}j^{12}j^{12}={1\over
2}\left(x^1p_2-x^2p_1+i\theta^1_\alpha\theta^2_\alpha \right)^2
\qquad ,
\]
in $I^\prime_2$ with an obvious physical meaning. In models of
Type III, $I_2$ contains the term $j^{12}p_2$ without the need to
add anything.

\subsubsection{SuperLiouville Models of Type IV:}
\begin{equation}
I_2^{(B)}= \frac{1}{2} p_1 p_1 +\frac{1}{2} \frac{\partial
W^{(a,b)}}{\partial x^1} \frac{\partial W^{(a,b)}}{\partial x^1}
\hspace{2cm} I_2^{(F)}=  i \frac{\partial^2 W^{(a,b)}}{\partial
x^1
\partial x^1} \theta_1^1 \theta_2^1 \label{eq:sinvIV}
\end{equation}

Like $I_1^{(F)}$, $I_2^{(F)}$ depends on $b$: the second invariant
in superLiouville models differs for different supersymmetric
extensions of the parent Liouville model.

\subsection{Other invariants}

From the pattern shown in the table we also conclude that the
generator of R-symmetry $S_2$ and their highest non-zero power -
$S_2^2$ in two dimensions - are invariants.
\begin{itemize}
\item The condition in {\it BOX 7} is not coupled to the rest of equations in table 1.
It sets the only independent component of $S_{ijkl}$ to be
constant, $S_{1212}=c$. Thus, as mentioned at the end of
sub-Section \S 2.2 , we check that
\[
S_3=-\frac{1}{2}S_2^2=\theta_1^1 \theta_1^2 \theta_2^1 \theta_2^2
\]
is a constant of motion, an invariant for all the ${\cal N}=2$
supersymmetric mechanical systems with two bosonic degrees of
freedom because no restriction on the superpotential has been
imposed.

\item If all the tensors vanish in the generic expansion of the invariant, except
the choice $F_{ij}=\delta_{ij}$, one immediately sees that
\[
S_2=i\theta_1^1 \theta_2^1+i\theta_1^2 \theta_2^2
\]
is a first integral in two dimensional ${\cal N}=2$ supersymmetric
mechanical systems.
\end{itemize}

\section{Supersolutions}
Given the \lq\lq even", $I_1=I_1^{(B)}+I_1^{(F)}$ ,
$I_2=I_2^{(B)}+I_2^{(F)}$ , $S_2$, and \lq\lq odd", $Q_\alpha$,
invariants of a superLiouville mechanical system, the
supertrajectories are constrained to satisfy:
\begin{eqnarray}
I_1(x^j,p_j,\theta^j_\alpha )=i_1 \qquad &,& \qquad
I_2(x^j,p_j,\theta^j_\alpha )=i_2 \label{eq:HJinv}
\\ Q_\alpha(x^j,p_j,\theta^j_\alpha)=q_\alpha \qquad &,& \qquad S_2(\theta^j_\alpha)=s_2 \qquad ,
\label{eq:fHJinv}
\end{eqnarray}
where $i_1$, $i_2$, $q_\alpha$, and $s_2$ are time-independent
arbitrary quantities. The system of equations
(\ref{eq:HJinv})-(\ref{eq:fHJinv}) relates bosonic, $x^j(t)$, to
fermionic, $\theta^j_\alpha$, variables; the $x^j(t)$ coordinates
cannot be ordinary functions; instead, they take values in the
even subalgebra ${\cal B}^e_L$ of the underlying Grassmann algebra
${\cal B}_L$.
\subsection{The Heumann-Manton method}
To solve the complicated system of equations $(\ref{eq:HJinv})$,
we propose the method envisaged by Heumann and Manton in
\cite{Manton1}. $L=4$ is sufficient for our purposes, so that the
identity and the real monomials
\[
\xi_A \qquad , \qquad \xi_{AB}=i\xi_A\xi_B \qquad , \qquad
\xi_{ABC}=i\xi_A\xi_B\xi_C \qquad , \qquad
\xi_{1234}=-\xi_1\xi_2\xi_3\xi_4 \qquad ,
\]
where $A,B,C=1,2,3,4$, provide a basis of ${\cal B}_4$. The key
idea is to expand $x^j(t)$, $p_j(t)$ and $\theta^j_\alpha (t)$ on
this basis:
\begin{eqnarray}
x^j(t)= x_o^j(t)+x_{AB}^j(t)\xi_{AB}+ x^j_{1234}(t)\xi_{1234}
\hspace{0.1cm} &,& \hspace{0.1cm}
p_j(t)=p_j^o(t)+p_j^{AB}(t)\xi_{AB}+p_j^{1234}(t)\xi_{1234}
\label{eq:bexp}
\\ \theta^j_\alpha(t)= \lambda^j_{\alpha
A}(t)\xi_A+\lambda^j_{\alpha ABC}(t)\xi_{ABC} \qquad &,&
\label{eq:fexp}
\end{eqnarray}
where $x_o^j(t)$, $p_j^o(t)$, ( the body ), $x^j_{AB}(t)$,
$p_j^{AB}(t)$, $x^j_{1234}(t)$, and $p_j^{1234}(t)$ are ordinary
function whereas $\lambda^j_{\alpha A}(t)$ and $\lambda^j_{\alpha
ABC}(t)$, are ordinary Majorana spinors. Of course, there is
antisymmetry in the $A,B,C$ indices of $x^j_{AB}$, $p_j^{AB}$ and
$\lambda_{\alpha ABC}^j$.

To facilitate the notation, we call $I_M^{(B)}=B_M(x^j,p_j)$,
$M=1,2$, and $I_M^{(F)}=F_M(x^j,p_j,\theta_\alpha^j)$ the terms of
the invariants without and with fermionic variables respectively .
The expansion of all the invariants
\begin{eqnarray*}
B_M(x^j,p_j)=B_M^o+B_M^{AB}\xi_{AB}+B_M^{1234}\xi_{1234} \quad &,&
\quad  F_M(x^j,p_j)=F_M^{AB}\xi_{AB}+F_M^{1234}\xi_{1234} \\
Q_\alpha(x^j,p_j,\theta^j_\alpha)=Q^A_\alpha\xi_A+Q_\alpha^{ABC}\xi_{ABC}
\qquad &,& \qquad S_2(\theta^j_\alpha
)=S_2^{AB}\xi_{AB}+S_2^{1234}\xi_{1234}\qquad ,
\end{eqnarray*}
together with a parallel expansion of the integration constants,
\begin{eqnarray*}
i_1=i_1^o+i_1^{AB}\xi_{AB}+i_1^{1234}\xi_{1234} \qquad &,& \qquad
i_2=i_2^o+i_2^{AB}\xi_{AB}+i_2^{1234}\xi_{1234} \\
q_\alpha=q_\alpha^A\xi_A+q_\alpha^{ABC}\xi_{ABC} \qquad &,&
s_2=s_2^{AB}\xi_{AB}+s_2^{1234}\xi_{1234} \qquad,
\end{eqnarray*}
permits a layer-by-layer writing of the (\ref{eq:HJinv}) system:
\begin{eqnarray}
B_1^o=i_1^o \qquad \qquad &,& \qquad \qquad B_2^o=i_2^o
\label{eq:bla}\\ B_1^{AB}+F_1^{AB}=i_1^{AB} \qquad \qquad &,&
\qquad \qquad B_2^{AB}+F_2^{AB}=i_2^{AB} \label{eq:fla} \\
B_1^{1234}+F_1^{1234}=i_1^{1234} \qquad \qquad &,& \qquad \qquad
B_2^{1234}+F_2^{1234}=i_2^{1234} \label{eq:sla} \quad .
\end{eqnarray}
Here, $i_M^o$, $i_M^{AB}$, $i_M^{1234}$, $q_\alpha^A $,
$q_\alpha^{ABC}$, $s_2^{AB}$, and $s_2^{1234}$ are real numbers
and a tedious calculation gives:
\[
B_M^o=B_M(x^j_o,p_j^o)
\]
\[
B_M^{AB}={\partial B_M\over \partial
x^k}(x^j_o,p_j^o)x^k_{AB}+{\partial B_M\over \partial
p_k}(x^j_o,p_j^o)p_k^{AB}
\]
\begin{eqnarray*}
B_M^{1234}&=&{\partial B_M\over \partial
x^k}(x^j_o,p_j^o)x^k_{1234}+{\partial B_M\over \partial
p_k}(x^j_o,p_j^o)p_j^{1234}+{1\over 2}{\partial^2 B_M\over\partial
x^k\partial x^l}(x^j_
o,p_j^o)\varepsilon_{ABCD}x^k_{AB}x^l_{CD}+\\&+&{1\over
2}\varepsilon_{ABCD}\left({\partial^2 B_M\over\partial x^k\partial
p_l}(x^j_o,p_j^o)x^k_{AB}p_l^{CD}+{1\over 2}{\partial^2
B_M\over\partial p_k\partial x^l}(x^j_
o,p_j^o)p_k^{AB}x^l_{CD}+{1\over 2}{\partial^2 B_M\over\partial
p_k\partial p_l}(x^j_ o,p_j^o)p_k^{AB}p_l^{CD}\right)
\end{eqnarray*}
\[
F_1^{AB}={\partial^2 W\over\partial x^k\partial
x^l}(x^j_o)\varepsilon_{ABCD}\lambda^k_{1C}\lambda^l_{2D}
\]
\[
F_1^{1234}=\varepsilon_{ABCD}\left({\partial^2 W\over\partial
x^k\partial
x^l}(x^j_o)(\lambda^k_{1A}\lambda^l_{2BCD}+\lambda^k_{1ABC}\lambda^l_{2D})+
{\partial^3 W\over\partial x^k\partial x^l\partial
x^m}(x^j_o)x^m_{AB}\lambda^k_
{1C}\lambda^l_{2D}\right)
\]
\begin{eqnarray*}
Q_\alpha^A&=&p_k^o\lambda^k_{\alpha
A}-\varepsilon_{\alpha\beta}{\partial W\over \partial
x^k}(x^j_o)\lambda^k_{\beta A}\\
Q_\alpha^{ABC}&=&p_k^o\lambda^k_{\alpha
ABC}-\varepsilon_{\alpha\beta}{\partial W\over \partial
x^k}(x^j_o)\lambda^k_{\beta
ABC}+\varepsilon_{ABCD}\left(p_k^{DE}\lambda^k_{\alpha
F}-\varepsilon_{\alpha\beta}{\partial^2 W\over \partial
x^k\partial x^l}(x^k_o)x^l_{DE}\lambda^k_{\beta F}\right)
\end{eqnarray*}
\[
S_2^{AB}=\varepsilon_{ABCD}\left(\lambda^1_{1C}\lambda^1_{2D}+\lambda^2_{1C}\lambda^2_{2D}\right)
\quad , \quad
S_2^{1234}=\varepsilon_{ABCD}(\lambda^1_{1A}\lambda^1_{2BCD}+\lambda^2_{1A}\lambda^2_{2BCD}
+\lambda^1_{1ABC}\lambda^1_{2D}+\lambda^2_{1ABC}\lambda^2_{2D})
\]
\vspace{0.5cm}

It is not possible to calculate the $F_2^{AB}$ and $F_2^{1234}$
components of the second invariant in a unified way because they
are different for different Types. Nevertheless, the first three
Types share a common structure, namely:
\[
F_2(x^j,p_j,\theta^j_\alpha)=il^{12}(x_o^j,p_j^o)\theta^1_\alpha\theta^2_\alpha+f^{kl}(x^j_o)
\theta^k_1\theta^l_2 \qquad ,
\]
where $f^{kl}(x^j_o)$ can be identified from (\ref{eq:sinvI}),
(\ref{eq:sinvII}), and (\ref{eq:sinvIII}) in each case. We thus
find:
\[
F_2^{AB}=\varepsilon_{ABCD}\left(l^{12}(x^j_o,p_j^o)\lambda^1_{\alpha
C}\lambda^2_{\alpha
D}-if^{kl}(x^j_o)\lambda^k_{1C}\lambda^l_{2D}\right)
\]
\begin{eqnarray*}
F_2^{1234}&=&
\varepsilon_{ABCD}\left[l^{12}(x^j_o,p_j^o)(\lambda^1_{\alpha
A}\lambda^2_{\alpha BCD}+\lambda^1_{\alpha ABC}\lambda^2_{\alpha
D})+({\partial l^{12}\over\partial x^k}(p_j^o)x^k_{AB}+{\partial
l^{12}\over \partial p_k}(x_o^j)p_k^{AB})\lambda^1_{\alpha
C}\lambda^2_{\alpha D} \right]+\\&+&
\varepsilon_{ABCD}\left[f^{kl}(x^j_o)(\lambda^k_{1 A}\lambda^l_{2
BCD}+\lambda^k_{1 ABC}\lambda^l_{2 D})-i{\partial f^{kl}\over
\partial x^m}(x_o^j)x^m_{AB}\lambda^k_{1 C}\lambda^l_{2 D} \right] \qquad .
\end{eqnarray*}
The equations for the basic layer (\ref{eq:bla}) ruling the
dynamics of the body of the system can be reduced to quadratures,
as in the original Liouville model. Except for Liouville models of
Type IV, this not longer holds for equations
(\ref{eq:fla})-(\ref{eq:sla}) in the first and second layers,
where the variables describing different degrees of freedom become
entangled. Nevertheless, we can be helped by considering equations
(\ref{eq:fHJinv}) provided by the fermionic invariants:
\begin{eqnarray}
Q_\alpha^A=q_\alpha^A \qquad \qquad &,& \qquad \qquad
Q_\alpha^{ABC}=q_\alpha^{ABC} \\ S_2^{AB}=s_2^{AB} \qquad \qquad
&,& \qquad \qquad S_2^{1234}=s_2^{1234}\qquad .
\end{eqnarray}

\subsection{Supersymmetry versus separability}
In Type IV models, the two degrees of freedom completely split in
equations (\ref{eq:bla}), (\ref{eq:fla}), (\ref{eq:sla}): we have
twice the solutions discussed in Reference \cite{Manton1}. We
shall now analyze the situation in the other three cases using the
coordinate system where the equations for the basic layer are
separable.
\begin{enumerate}
\item Type I

\bigskip

Using elliptic coordinates, the invariants of the Type I systems
are:
\[
\xi^*B_1={1\over 2(u^2-v^2)}\left[(u^2-c^2)(p_u^2+\left({dF\over
du} \right)^2)+(c^2-v^2)(p_v^2+\left({dG\over dv}
\right)^2)\right]
\]
\begin{eqnarray*}
\xi^*F_1&=& i\left[{d^2F\over du^2}+{u(c^2-v^2)\over
(u^2-v^2)(u^2-c^2)}{dF\over du}-{v(c^2-v^2)\over
(u^2-v^2)(u^2-c^2)}{dG\over
dv}\right]\vartheta_1^u\vartheta^u_2\\\\&+&i\left[{v\over
u^2-v^2}{dF\over du}+{u\over u^2-v^2}{dG\over
dv}\right]\left(\vartheta^u_1\vartheta^v_2+\vartheta^v_1\vartheta^u_2\right)\\
&+&i\left[{d^2G\over dv^2}+{u(u^2-c^2)\over
(u^2-v^2)(c^2-v^2)}{dF\over du}-{v(u^2-c^2)\over
(u^2-v^2)(c^2-v^2)}{dG\over dv}\right]\vartheta_1^v\vartheta^v_2
\end{eqnarray*}
\[
\xi^*B_2={(u^2-c^2)(c^2-v^2)\over
2(u^2-v^2)}\left\{\left[p_v^2+\left({dG\over dv}
\right)^2\right]-\left[p_u^2+\left({dF\over du}
\right)^2\right]\right\}
\]
\begin{eqnarray*}
\xi^*F_2&=&
i(vp_u-up_v)\vartheta_\alpha^u\vartheta_\alpha^v\\&-&i(c^2-v^2)\left[{d^2F\over
du^2}+{u(c^2-v^2)\over (u^2-v^2)(u^2-c^2)}{dF\over du}-{v\over
u^2-c^2}\left({c^2-v^2\over u^2-v^2} -1\right){dG\over
dv}\right]\vartheta_1^u\vartheta^u_2\\\\&+&i\left[{u(c^2-v^2)\over
u^2-v^2}{dG\over dv}+{v(u^2-c^2)\over u^2-v^2}{dF\over
du}\right]\left(\vartheta^u_1\vartheta^v_2+\vartheta^v_1\vartheta^u_2\right)\\
&+&i(u^2-c^2)\left[{d^2G\over dv^2}+{u\over
c^2-v^2}\left({u^2-c^2\over u^2-v^2}-1\right){dF\over
du}-{v(u^2-c^2)\over (u^2-v^2)(c^2-v^2)}{dG\over
dv}\right]\vartheta_1^v\vartheta^v_2
\end{eqnarray*}
\[
\xi^*Q_\alpha=p_u\vartheta_\alpha^u-\varepsilon_{\alpha\beta}{dF\over
du}\vartheta_\beta^u
+p_v\vartheta_\alpha^v-\varepsilon_{\alpha\beta}{dG\over
dv}\vartheta_\beta^v \qquad , \qquad
\xi^*S_2=i(u^2-v^2)({\vartheta^u_1\vartheta^u_2\over u^2-c^2}
+{\vartheta^v_1\vartheta_2^v\over c^2-v^2})
\]

\item Type III

\bigskip

In parabolic coordinates the, invariants of the Type III systems
read:
\[
\zeta^*B_1={1\over 2(u^2+v^2)}\left[p_u^2+\left({dF\over du}
\right)^2+p_v^2+\left({dG\over dv} \right)^2\right]
\]
\begin{eqnarray*}
\zeta^*F_1&=& i\left[{d^2F\over du^2}-{u\over u^2+v^2}{dF\over
du}+{v\over u^2+v^2}{dG\over
dv}\right]\vartheta_1^u\vartheta^u_2\\\\&-&i\left[{u\over
u^2+v^2}{dG\over dv}+{v\over u^2+v^2}{dF\over
du}\right]\left(\vartheta^u_1\vartheta^v_2+\vartheta^v_1\vartheta^u_2\right)\\
&+&i\left[{d^2G\over dv^2}+{u\over u^2+v^2}{dF\over du}-{v\over
u^2+v^2}{dG\over dv}\right]\vartheta_1^v\vartheta^v_2
\end{eqnarray*}
\[
\zeta^*B_2={1\over 2(u^2+v^2)}\left\{u^2\left[p_v^2+\left({dG\over
dv} \right)^2\right]-v^2\left[p_u^2+\left({dF\over du}
\right)^2\right]\right\}
\]
\begin{eqnarray*}
\zeta^*F_2&=&
-i(vp_u+up_v)\vartheta_\alpha^u\vartheta_\alpha^v-iv^2\left[{d^2F\over
du^2}-{u\over u^2+v^2}{dF\over du}+\left({v\over
u^2+v^2}-1\right){dG\over
dv}\right]\vartheta_1^u\vartheta^u_2\\\\&-&iuv\left[{u\over
u^2+v^2}{dG\over dv}-{v\over u^2+v^2}{dF\over
du}\right]\left(\vartheta^u_1\vartheta^v_2+\vartheta^v_1\vartheta^u_2\right)\\
&+&iu^2\left[{d^2G\over dv^2}+\left({u\over
u^2+v^2}-1\right){dF\over du}-{v\over u^2+v^2}{dG\over
dv}\right]\vartheta_1^v\vartheta^v_2
\end{eqnarray*}
\[
\zeta^*Q_\alpha=p_u\vartheta_\alpha^u-\varepsilon_{\alpha\beta}{dF\over
du}\vartheta_\beta^u
+p_v\vartheta_\alpha^v-\varepsilon_{\alpha\beta}{dG\over
dv}\vartheta_\beta^v \qquad , \qquad
\zeta^*S_2=i(u^2+v^2)(\vartheta^u_1\vartheta^u_2+\vartheta^v_1\vartheta_2^v)
\]
Use of the expansions
\[
u(t)=u_o(t)+u_{AB}(t)\xi_{AB}+u_{1234}(t)\xi_{1234}\quad , \quad
v(t)=v_o(t)+v_{AB}(t)\xi_{AB}+v_{1234}(t)\xi_{1234}
\]
\[
p_u(t)=p_u^o(t)+p_u^{AB}(t)\xi_{AB}+p_u^{1234}(t)\xi_{1234}\quad ,
\quad p_v(t)=p_v^o(t)+p_v^{AB}(t)\xi_{AB}+p_v^{1234}(t)\xi_{1234}
\]
\[
\vartheta_\alpha^u(t)=\lambda_{\alpha A}^u(t)\xi_A+\lambda_{\alpha
ABC}^u(t)\xi_{ABC} \qquad , \qquad
\vartheta_\alpha^v(t)=\lambda_{\alpha A}^v(t)\xi_A+\lambda_{\alpha
ABC}^v(t)\xi_{ABC}
\]

in the system of equations ruling the dynamics of the Type I and
Type III systems,
\begin{eqnarray}
\xi^*I_1&=&i_1 \quad , \quad \xi^*I_2=i_2 \quad , \quad
\xi^*Q_\alpha=q_\alpha \quad , \quad \xi^*S_2=s_2 \label{eq:hjI}\\
\zeta^*I_1&=&i_1 \quad , \quad \zeta^*I_2=i_2 \quad , \quad
\zeta^*Q_\alpha=q_\alpha \quad , \quad \zeta^*S_2=s_2 \qquad ,
\label{eq:hjII}
\end{eqnarray}
leads to a layer-by-layer solution of the problem.

In the basic layer, with no Grassman variables at all, the
dynamics of the $u$ and $v$ variables are completely independent
with respect to each other and a solution by quadratures is at
hand. One can easily check that this is not the case in the second
and third layer: owing to the Yukawa terms, the dynamics of the
$u$ and $v$ variables are no longer independent in the
supersymmetric extension of this kind of system.

\item Type II

\bigskip

The behaviour of Type II models is identical to the behaviour
described above for Type I and III systems. We thus merely list
the invariants of this Type of model for completeness using polar
coordinates:
\[
\rho^*B_1={1\over 2}\left[p_\rho^2+\left({dF\over
d\rho}\right)^2\right]+{1\over
2\rho^2}\left[p_\chi^2+\left({dG\over d\chi}\right)^2\right]
\]
\[
\rho^*F_1=i{d^2F\over
d\rho^2}\vartheta_1^\rho\vartheta_2^\rho+i\left({d^2G\over
d\chi^2}+\rho{dF\over
d\rho}\right)\vartheta_1^\chi\vartheta_2^\chi+{1\over
\rho}{dG\over
d\chi}(\vartheta_1^\rho\vartheta_2^\chi+\vartheta_1^\chi\vartheta_2^\rho
)
\]
\[
\rho^*B_2={1\over 2}\left\{p_\chi^2+\left({dG\over
d\chi}\right)^2\right\} \qquad , \qquad
\rho^*F_2=-i\rho^2p_\chi\vartheta_\alpha^\rho\vartheta_\alpha^\chi+i\rho^2{d^2G\over
d\chi^2}\vartheta_1^\chi\vartheta_2^\chi
\]
\[
\rho^*Q_\alpha=p_\rho\vartheta_\alpha^\rho-\varepsilon_{\alpha\beta}{dF\over
d\rho}\vartheta_\beta^\rho+p_\chi\vartheta_\alpha^\chi-\varepsilon_{\alpha\beta}{dG\over
d\chi}\vartheta_\beta^\chi \qquad , \qquad
\rho^*S_2=i(\vartheta_1^\rho\vartheta_2^\rho+\rho^2\vartheta_1^\chi\vartheta_2^\chi
)
\]
The expansions
\[
\rho(t)=\rho_o(t)+\rho_{AB}(t)\xi_{AB}+\rho_{1234}(t)\xi_{1234}\quad
, \quad
\chi(t)=\chi_o(t)+\chi_{AB}(t)\xi_{AB}+\chi_{1234}(t)\xi_{1234}
\]
\[
p_\rho(t)=p_\rho^o(t)+p_\rho^{AB}(t)\xi_{AB}+p_\rho^{1234}(t)\xi_{1234}\quad
, \quad
p_\chi(t)=p_\chi^o(t)+p_\chi^{AB}(t)\xi_{AB}+p_\chi^{1234}(t)\xi_{1234}
\]
\[
\vartheta_\alpha^\rho(t)=\lambda_{\alpha
A}^\rho(t)\xi_A+\lambda_{\alpha ABC}^\rho(t)\xi_{ABC} \qquad ,
\qquad  \vartheta_\alpha^\chi(t)=\lambda_{\alpha
A}^\chi(t)\xi_A+\lambda_{\alpha ABC}^\chi(t)\xi_{ABC}
\]
allow to organize the dynamics in a layer-by-layer structure.

\end{enumerate}

\section*{Acknowledgements}

A. Alonso-Izquierdo gratefully acknowledges the support of the
spanish Secretaria de Estado de Educacion y Universidades
co-financed by the Fondo Social Europeo.

\end{document}